%% file: mnras_template.tex
\documentclass[usenatbib]{mnras}

\usepackage[modulo]{lineno}
\setpagewiselinenumbers
\modulolinenumbers[1]

\usepackage[T1]{fontenc}

\usepackage{mathrsfs}
\DeclareRobustCommand{\VAN}[3]{#2}
\let\VANthebibliography\thebibliography
\def\thebibliography{\DeclareRobustCommand{\VAN}[3]{##3}\VANthebibliography}

\usepackage{graphicx}	
\usepackage{amsmath}	
\usepackage{amssymb}	
\usepackage{multirow}
\usepackage{subcaption}

\title[Gas Perturbations in Hot Smooth Atmospheres]{X-ray Surface Brightness Fluctuations in Smooth Galaxy Cluster Atmospheres}

\author[M. Li et al.]{
Muzi Li,$^{1,2}$\thanks{E-mail: muzi.li@uwaterloo.ca}
B. R. McNamara,$^{1,2,3}$
Irina Zhuravleva$^{4}$\\
$^{1}$Department of Physics and Astronomy, University of Waterloo, 200 University Avenue West, Waterloo, ON N2L 3G1, Canada\\
$^{2}$Waterloo Center for Astrophysics, University of Waterloo, 200 University Avenue West, Waterloo, ON N2L 3G1, Canada\\
$^{3}$Perimeter Institute for Theoretical Physics, Waterloo, ON N2L 2Y5, Canada\\
$^{4}$Department of Astronomy \& Astrophysics, University of Chicago, 5640 S Ellis Avenue, Chicago, IL 60637, USA
}

\date{Accepted in Jan 2025 by MNRAS}

\pubyear{2015}

\begin{document}

\label{firstpage}
\pagerange{\pageref{firstpage}--\pageref{lastpage}}
\maketitle

\begin{abstract}
We measure surface brightness fluctuations in Chandra X-ray images of the cores of the galaxy clusters Abell 2029, Abell 2151, Abell 2107, RBS0533, and RBS0540. Their relatively structureless X-ray atmospheres exhibit the thermodynamic properties of cool cores including short central cooling times and low entropy. However, unlike typical cool-core clusters, molecular gas, star formation, and bubbles associated with radio jets are faint or absent near their central galaxies. Four clusters show typical gas density fluctuation amplitudes of $\sim$ 10 per cent on the scales probed, apart from RBS0540, which exhibits lower amplitudes, suggesting that its gas is mildly disturbed. Under the assumption that gas density fluctuations are indicative of random gas velocities, we estimate scale-dependent velocity amplitudes of gas motions across all studied clusters, which range from 100 $\rm km~s^{-1}$ to 200 $\rm km~s^{-1}$ in Abell 2029, Abell 2151, and Abell 2107. These velocity estimates are comparable to the atmospheric velocity dispersion in the Perseus cluster measured by the Hitomi X-ray Observatory. The turbulent heating rates implied by our measurements are of the same order as the radiative cooling rates. Our results suggest that atmospheric sloshing and perhaps turbulent motion may aid radio jets in stabilizing atmospheric cooling.

\end{abstract}



\input{Cpt1_Introduction}

\input{Cpt2_Data_reduction}

\input{Cpt3_Power_spectral}

\input{Cpt4_Results}
\input{Cpt5_Discussion}

\input{Cpt6_Conclusion}

\input{Acknowledgements}

\input{data_availability}



\bibliographystyle{mnras}
\bibliography{reference} 




\appendix


\bsp	
\label{lastpage}
\end{document}

%% file: Cpt1_Introduction.tex
\section{Introduction}
\label{sec:intro}
In the absence of reheating, the hot intracluster medium (ICM) would cool radiatively leading to cooling into molecular clouds and star formation at rates ranging from hundreds to thousands of solar masses per year \citep[e.g.][]{fabian1994cooling}, violating observation \citep*[]{edge2001molecular, rafferty2008regulation,pulido2018origin}. A variety of heat sources have been proposed to suppress cooling. The most persistent and capable heat source is the energy released from the central supermassive black hole traced by radio jets and X-ray bubbles embedded in hot atmospheres \citep[e.g.][]{fabian2012observational,mcnamara2007heating,churazov2000asymmetric}. 
Chandra observations have shown that the energy released by buoyantly rising X-ray bubbles is able to balance atmospheric cooling over five decades of X-ray cooling luminosity \citep[e.g.][]{mcnamara2000chandra,rafferty2006feedback}. However, how and how much jet energy is converted to atmospheric heating remains uncertain.  Jet energy is transported throughout the atmosphere by a combination of weak shocks \citep[e.g.][]{heinz1998x,fabian2006very,markevitch2007shocks,randall2010shocks}, sound waves \citep[e.g.][]{sanders2007deeper,sternberg2009sound,nulsen2013agn}, and turbulence generated in wakes of the X-ray bubbles \citep*[]{mcnamara2007heating,graham2008weak}. Other mechanisms include heating from cosmic rays \citep[e.g.][]{loewenstein1991cosmic,guo2008feedback,pfrommer2013toward}, gravity waves~\citep*[]{zhang2018generation}, and the release of gravitational potential energy by slow gas displacements \citep*[]{hillel2016heating,hillel2017hitomi}. Regardless of sources that drive gas perturbations, the measurements based on power spectra analysis of X-ray surface brightness (SB) fluctuations suggest that dissipation of subsonic turbulence is, on average, sufficient enough to compensate for radiatively cooling loss and reheat ICM in cluster cores \citep*[e.g.][]{zhuravleva2014turbulent,zhuravleva2018gas}.

As the cold, centrally accreting gas fuels the nuclear activities that suppress cooling is an essential part of the feedback loop, molecular clouds, bright nebular line emission and enhanced star formation are commonly observed in central cluster galaxies when the central entropy and cooling timescales fall below $\rm \sim 30\ keV\ cm^2$ and $\rm \sim 1\ Gyr$, respectively \citep[e.g.][]{cavagnolo2008entropy,rafferty2008regulation}. The cold gas is expected to condense out of hot atmospheres when the ratio of cooling timescale to the free-fall timescale lies below the unity, $\rm t_{cool}/t_{ff} \lesssim 1$ \citep[e.g.][]{nulsen1986thermal,balbus1989theory,mccourt2011can,voit2015cooling}. Nevertheless, the cooling time of the hot ambient medium pervading a massive galaxy does not drop much below ten times the free-fall timescale at any radius, even in the cores of clusters \citep[e.g.][]{hogan2017onset,pulido2018origin,babyk2019origins,voit2015cooling}. This again indicates that atmospheres are globally thermally stable. However, something must trigger thermal instability locally so that multi-phase condensation can proceed \citep[e.g.][]{pizzolato2005nature,pizzolato2010solving,mccourt2012thermal,mcnamara2016mechanism,gaspari2018shaken}.

Abell 2029 is the archetype in a handful of exceptions. The radiative cooling time in its inner 10 kpc falls below $\rm 5\times10^8~ yr$, dropping to $\rm 2\times 10^8 ~yr$ within the central 3 kpc. Its gas density rises to $0.1\rm ~cm^{-3}$ yielding one of the known brightest atmospheric cusps of thermal X-ray emission. The ICM should be cooling into molecular clouds at a rate of several hundred solar mass per year \citep[]{sarazin1992x}. However, the optical observations are inconsistent with the relaxed cooling-flow picture of star formation, and nebular emission is absent at the levels seen in other clusters and groups with short central cooling time. Apart from a prominent swirl in its hot atmosphere associated with a cold front \citep[]{clarke2004complex,paterno2013deep}, a feature common to many clusters with burgeoning central galaxies such as Perseus \citep[e.g.][]{boehringer1993rosat,fabian2003deep}, it reveals no X-ray cavities inflated by its central radio source. 

\citet{mcnamara2016mechanism} proposed that cold gas condenses from low entropy gas that is lifted outward from cluster cores to an altitude where its cooling time becomes shorter than its free-fall timescale, such that $\rm t_{cool}/t_{ff}\lesssim 1$, by buoyantly rising X-ray bubbles. \citet{martz2020thermally} studied five cool core clusters: Abell 2029, Abell 2107, Abell 2151, RBS0533, and RBS0540 with relatively short central cooling times that show no evidence for cooling traced by nebular gas and star formation. Their analysis suggested that the absence of X-ray bubbles prevents the effective lifting of low-entropy gas to an altitude where the ratio of cooling time to free-fall time approaches unity. Thus they may temporarily be thermally stable. 

The objective of this paper is to understand better why the five galaxy clusters, Abell 2029, Abell 2151, Abell 2107, RBS0533, and RBS0540, those lacking strong signatures of radio jet interaction, exhibit smooth atmospheres with short cooling times but do not appear to be cooling at appreciable rates.

Despite the absence of definitive evidence for X-ray cavities, the dissipation rate of gas turbulence, $\rm Q_{turb}$, in the core of Abell 2029 generates sufficient heating to locally counteract the radiative cooling rate, $\rm Q_{cool}$, although there remains considerable uncertainty \citep[]{zhuravleva2018gas}. It suggests that turbulent dissipation driven by other processes may play an important role in preventing cooling in hot smooth atmospheres. For instance, X-ray observations frequently reveal spiral-like structures known as ‘cold fronts’ in the ICM of numerous clusters. These cold fronts, marked by sharp edges in temperature and density, can be seen in clusters such as Virgo \citep[]{roediger2011gas}, Perseus \citep[]{fabian2003deep,fabian2006very}, and Abell 2204 \citep[]{chen2017gas}. The formation of these cold fronts can result from mergers with subhalos, which cause the cold gas to ‘slosh’ out from the cores of clusters \citep[]{markevitch2007shocks}. The spiral features seen in X-ray residual images of two smooth atmosphere clusters in our sample, Abell 2029, and Abell 2151, are evidence of interaction and may indicate their potential to produce sufficient turbulence \citep[e.g][]{zuhone2011sloshing,walker2018fraction} to prevent ICM from cooling, at least temporarily. Moreover, gas can be disturbed by jet-driven shock \citep[]{friedman2012all} and by galaxy motions \citep[]{gu2013search}.

Using Chandra X-ray observations, we performed a new statistical analysis \citep[]{arevalo2012mexican,zhuravleva2014turbulent,zhuravleva2014relation} of X-ray surface brightness fluctuations in the cool-cores of these five smooth atmosphere clusters. Since this technique takes advantage of the weak temperature-dependence of the atmospheric emissivity in the soft energy band (0.5 - 3.5 keV is used in this work), the amplitudes emissivity fluctuations can be interpreted as equivalent to the amplitudes of density fluctuations measured over a range of spatial scales. \citep[e.g.][]{forman2007filaments}. At higher energies (3.5 - 7 keV), emissivity depends strongly on temperature; therefore, the relative amplitudes of SB fluctuations (cross-spectrum) in the soft and hard band are sensitive to the atmosphere's effective equation of state, which enables us to understand the nature of gas perturbations \citep[]{zhuravleva2016nature,zhuravleva2018gas,10.1093/mnras/stw2044}. Although high spatial-resolution Chandra observations fit in with such analysis well, pre-existing data of clusters studied here in the archive contains too few counts to yield reliable measurements on small scales where Poisson noise dominates. This lack of counts is amplified in the high energy band; hence effective equation of state of gas perturbations won't be measured in this work. Longer Chandra observations are required. To further investigate if their atmospheres in the core regions can be stabilized by turbulent dissipation, even though they lack AGN-inflated bubbles, we derive the velocity power spectra of gas motions and estimate the turbulent heating rates.

The structure of this paper is organized as follows: In Section \ref{sec:sec2}, we describe the selected cluster samples with smooth atmospheres and outline the main steps of data reduction. Section \ref{sec: power_spectrum} details the power spectrum analysis methodology. The results of the perturbation analysis are presented in Section \ref{sec:results} and discussed in Section \ref{sec: discussion}. Finally, we provide our conclusions in Section \ref{sec: conclusion}.  We here adopt $\rm \Lambda CDM$ cosmology with $\Omega_m=0.3$, $\Omega_\Lambda=0.7$ and $h=0.72$ throughout this work.

%% file: Cpt2_Data_reduction.tex
\section{Smooth Atmosphere Clusters and X-ray Images}

\label{sec:sec2}

\begin{figure*}
\centering

	\includegraphics[width=0.85\linewidth]{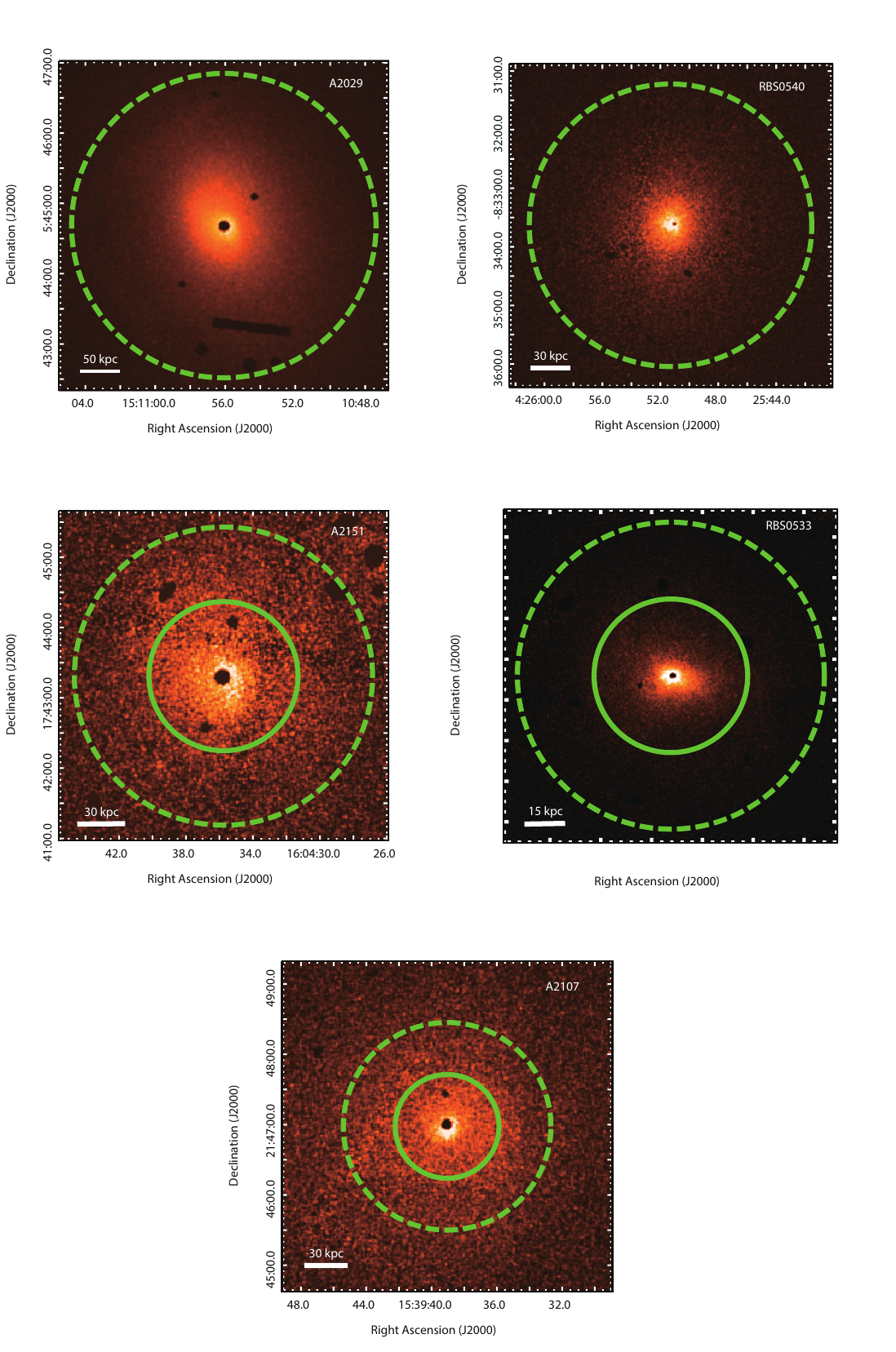}
    \caption{The background-subtracted, exposure-corrected Chandra images of our targets in the $0.5 - 3.5$ keV band. The cool-core radius is indicated with dashed green circles, and the inner half of the cool-core radius is marked in the green solid line. The bright point sources and the central AGNs have been masked out. The images are smoothed with a 3'' Gaussian for display purposes.}
    \label{fig:chandra_figure_spoiler}
\end{figure*}

\begin{figure*}
\centering
	\includegraphics[width=16cm,height=7cm]{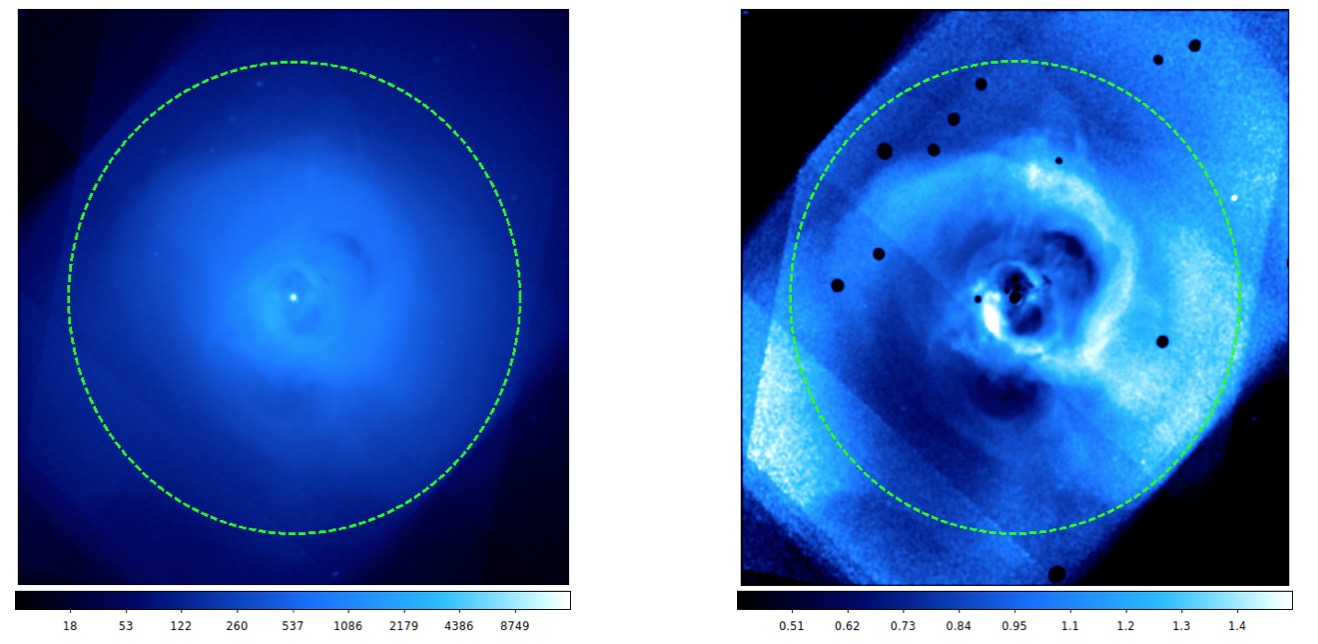}
    \caption{Left: Background-subtracted, exposure-corrected mosaic Chandra images of the Perseus cluster in the $0.5 - 4$ keV band. The dashed green circle indicates the inner region within half cool core radius, denoted as $0 < r < r_{\text{cool}}/2$. Each image is smoothed with a 3'' Gaussian for display purposes.
    Right: Residual X-ray image of the Perseus cluster, obtained by dividing the initial image by the best-fitting spherically symmetric $\beta$-model. The bright point sources and the central AGNs have been masked out.}
    \label{fig:chandra_figure_perseus}
\end{figure*}

\subsection{Sample Selection}

We focus on the X-ray surface brightness analysis in the cool cores of galaxy clusters Abell 2029, Abell 2107, Abell 2151, RBS0533, and RBS0540. These clusters were first identified as outliers from the \texttt{ACCEPT} database, based on the upper limits on H$\alpha$ luminosity as suggested by \citet{cavagnolo2008entropy}. Despite the absence of $\rm H\alpha$ emission and star formation at levels observed in other systems with similar thermodynamic properties, their central cooling times and central entropy fall below the thresholds of $\rm 10^9~yr$ and $30\ \rm keV\ cm^{2}$, respectively. Our small sample spans a range of environments, from groups to rich clusters, with central densities and central pressures lying between $0.01-0.1\ \rm cm^{-3}$ and $10^{-9}-10^{-10}\ \rm erg\ cm^{-3}$. Upon examining Chandra X-ray images of these five clusters, all atmospheres lack X-ray cavities, apart from RBS0533, which may have a potential cavity southeast of its center but with low significance. The relaxed atmospheres of Abell 2029 and Abell 2151 display spiral structures probably introduced by the `slosh' motion of core gas arising from small infalling groups or subclusters. In this work, we study why the clusters in our sample remain thermally stable despite the absence of X-ray bubbles, which are typically associated with energetic central active galactic nucleus (AGN) activities. We explore whether contributions from other sources of turbulence might be capable of compensating for radiative cooling. 

\label{sec:sample_selection}

\begin{table*}
\begin{tabular}{llcccccc}
\hline
Cluster & \multicolumn{1}{c}{ObsIDs}     & Redshift & \multicolumn{2}{c}{BCG Position} & Cleaned $\rm T_{exp}$ & $\rm N_{H}$ & $\rm r_{cool}$ \\
        & \multicolumn{1}{c}{}           &          & RA (J2000)      & DEC J(2000)    & ks                      & $\rm 10^{22}cm^{-2}$ & kpc            \\
(a)     & \multicolumn{1}{c}{(b)}        & (c)      & (d)             & (e)            & (f)                     & (g)              & (h)            \\ \hline
RBS0533 & 3186, 3187, 5800, 5801         & 0.0123   & \multicolumn{1}{c}{4:19:38.105}     & \multicolumn{1}{c}{+2:24:35.54}    & 107.9                   & 0.102            & 58             \\
Abell 2029   & 891, 4977, 6101                & 0.0773   & \multicolumn{1}{c}{15:10:56.077}    & \multicolumn{1}{c}{+05:44:41.05}   & 103.3                   & 0.033            & 180            \\
Abell 2151   & 4996, 19592, 20086, 20087      & 0.0366   & \multicolumn{1}{c}{16:04:35.758}    & \multicolumn{1}{c}{+17:43:18.54}   & 80.2                    & 0.033            & 94             \\
RBS0540 & 4183, 19593, 20862, 20863      & 0.0397   & \multicolumn{1}{c}{4:25:51.300}     & \multicolumn{1}{c}{-8:33:38.00}    & 61.6                    & 0.079            & 115            \\
Abell 2107   & 4960,24350,24351,26152,        & 0.0410   & \multicolumn{1}{c}{15:39:39.043}    & \multicolumn{1}{c}{+21:46:58.55}   & 140.0                   & 0.045            & 71             \\
        & 26153,26443, 23851             &          &                 &                &                         &                  &                \\
Perseus & 3209, 4289, 4946, 4947 - 4953, & 0.0176   & \multicolumn{1}{c}{03:19:48.160}    &\multicolumn{1}{c}{ +2:16:1.00}     & 1335.5                  & 0.136            & 174            \\
        & 12025, 12033, 12036, 12037     &          &                 &                &                         &                  &               \\ \hline
\end{tabular}
\caption{Clusters and Chandra observations in our analysis: (a). Cluster name, (b). Observations used in this analysis, (c). Redshift, (d). Right Ascension (J2000), (e). Declination (J2000), (f). Cleaned exposure after data reduction, (g). Column density, and (h). Cool-core radius.}
\label{tab:observations}
\end{table*}

To explore the properties of gas perturbation natures of structureless atmosphere clusters in our sample, we also present the analysis within the half-cool-core region ($\sim 4.1'$ / 436 kpc) of Perseus cluster as a point of reference, using a total of 1.4 Ms of deep Chandra observations. The Chandra observations of our sample clusters and Perseus cluster are detailed in Table~\ref{tab:observations}. The Perseus cluster is a bright, nearby galaxy cluster where the central cooling time is an order of magnitude lower than the Hubble time, and the central entropy also lies below $30\ \rm keV\ cm^{2}$. Compared with our samples with smooth atmospheres, X-ray observations reveal AGN-inflated bubbles of relativistic plasma in the core of Perseus, surrounded by weak shocks \citep[e.g][]{boehringer1993rosat,fabian2003deep,fabian2006very}. It indicates that the gas in the central region of the Perseus cluster is strongly disturbed, and gas turbulent motions might be triggered during the inflation and buoyant rising of those bubbles. \citet{zhuravleva2014turbulent} measured the amplitude of gas perturbations in the core of Perseus from its X-ray surface brightness fluctuation power spectrum. They found that central AGN activities might play a dominant role in driving gas turbulent motions in the cool core of Perseus, and the heating rate due to the dissipation of gas turbulence should be sufficient to balance radiative loss, maintaining an approximately stable state in the ICM.


\subsection{Data Reduction}
\label{sec:data_reduction}
The analysis in this work is based on archival Chandra X-ray data utilizing the Advanced CCD Imaging Spectrometer (ACIS) detectors ACIS-I and ACIS-S. Observations used in this analysis are summarized in Table~\ref{tab:observations}. All have been reprocessed following  \citet{vikhlinin2005chandra}. The observations were reduced using \texttt{CIAO} version 4.12, \texttt{CALDB} version 4.9.2. Light curves were extracted from level-2 event files above 10 keV using the \texttt{lc\_clean} script to eliminate intervals affected by background flares. Blank sky files were normalized to count rates in the $10 - 12$ keV range and projected to the corresponding positions. Exposure maps were calculated to correct for varying exposure coverage caused by chip gaps and vignetting. Multiple observations were projected onto the one with the longest exposure and summed into a background-subtracted, exposure-corrected image. Point sources, including the central AGN, were identified using the \texttt{wavdetect} tool and by inspection and were removed accounting for the shape of the Point Spread Function (PSF).  

The X-ray emissivity, $\Lambda(T)$ in the low energy band ($0.5 - 3.5$ keV in this work), is largely independent of the gas temperature when higher than $\sim 3$ keV. Therefore, surface brightness fluctuations shown in this band are used to infer ICM density fluctuations. Three clusters have a volume-weighted mean temperature lying below 3 keV within the region of interest: RBS0533 at $\sim 1.2$ keV, Abell 2151 at $\sim 2.2$ keV, and RBS0540 $\sim$ 2.8 keV. To validate the temperature independence in each target, we employed a narrower and softer energy band of $0.5 - 1.5$ keV to measure perturbations and compared these with results from the $0.5 - 3.5$ keV band. The discrepancies between the two energy band measurements are less than 6\%. The narrower energy band yields fewer counts which limits the smallest scales we can probe and amplifies the uncertainties primarily due to Poison noise. Therefore, X-ray images were prepared in the $0.5 - 3.5$ keV band given the minimal temperature dependence in this `soft' energy band and its higher photon count contribution.

Figure~\ref{fig:chandra_figure_spoiler} displays the reprocessed Chandra images of the sample cluster cores with smooth atmospheres in the $0.5 - 3.5$ keV band. Green dashed circles indicate the areas of the cool cores. The cool-core radius is summarized in column (h) of Table~\ref{tab:observations} for each target. The measurement precision within the cool cores of RBS0533, Abell 2151, and Abell 2107 is limited by their relatively low surface brightness or the existing Chandra exposure. We thus reduced our regions of interest to the inner half of cool-cores in these three objects, which are marked as solid green circles in Figure~\ref{fig:chandra_figure_spoiler}. The deep Chandra X-ray mosaic of the Perseus core is created in 0.5 - 4 keV and is presented in the left panel of Figure~\ref{fig:chandra_figure_perseus}. The green dashed circle shows the central region we studied, within the half cool-core radius ($\sim 4.1'/436$ kpc). 


\subsubsection{Deprojected Thermodynamic Profiles}
\label{sec:spectral_fitting}
Spectra were extracted from concentric circular annuli centered on the brightest cluster galaxy (BCG) within the 0.5 - 7 keV energy band, with the coordinates of the BCGs detailed in Table~\ref{tab:observations}. The central annulus contains at least 3000 counts. The emission observed at any location contains superposed emission from hotter, overlying layers. To measure and remove the overlying emission, the spectra were deprojected using \texttt{DSDEPROJ} \citep {russell2008direct,sanders2007deeper}.  Spectra were extracted separately from each observation and grouped to ensure a minimum of 30 counts per energy bin. To ensure accurate deprojected temperature measurements the net counts in subsequent annuli were set to be 1.2 times the count of the preceding one.

Thermodynamic properties were derived from spectra using an absorbed thermal model \texttt{Phabs*(apec)} built in \texttt{XSPEC}. The initial values for the galactic hydrogen column density used for fitting were based on the measurements provided by \citet{kalberla2005leiden}, as listed in column (g) of Table~\ref{tab:observations}. This parameter was then allowed to vary freely within each spectral bin. The heavy element abundance was treated as a free parameter.

Using the radial profiles of temperature $\rm T$ and electron number density $\rm n_e$, we calculated the entropy index of the ICM as $\rm K=k_B T n_e^{-2/3}$, where $\rm k_B$ is the Boltzmann constant. The cooling timescale was calculated using:

\begin{equation}
    {\rm t_{cool}=\frac{3}{2}\frac{(n_e+n_i) k_B T}{n_e n_i \Lambda(T)}},
    \label{eq:cooling_time}
\end{equation}
where $\rm n_i=(\xi-1)n_e$ is the ion number density with $\xi=1.912$ for fully ionized gas. The normalized cooling function, $\Lambda(T)$,  assuming $Z_\odot =0.3$ was derived from the tabulations by \citet{sutherland1993cooling}. The cool-core radius in this work denoted as $\rm r_{cool}$, is defined as the distance from the center within which the cooling timescale of the ICM is shorter than the Hubble time, which is $\sim 1.36$ $\rm \times 10^{10}~yr$ according to the cosmological parameters we adopted.

Assuming identical ion and electron temperatures and an ideal monatomic gas, the sound speed $\rm c_s$ is given by:

\begin{equation}
    c_s=\sqrt{\gamma\frac{k_B T}{\mu m_p}},
    \label{eq:sound_speed}
\end{equation}
where $\gamma=5/3$ is the index for monatomic ideal gas, and $\mu=0.61$ and $m_p$ are the mean particle mass and the proton mass, respectively. 

\begin{figure*}
    \begin{subfigure}{.45\textwidth}
		\centering
		\includegraphics[width=\linewidth]{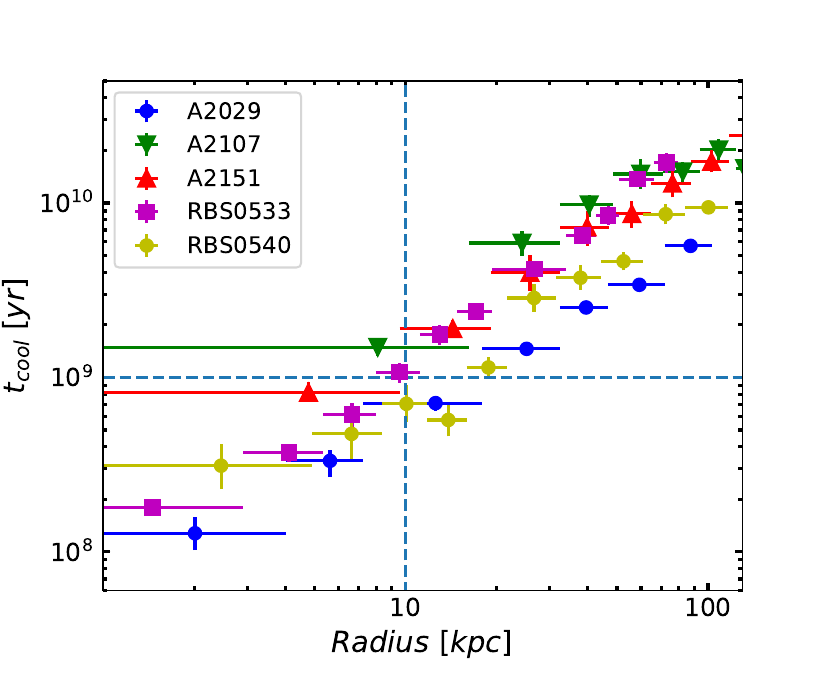}
		\caption{}
	\end{subfigure}
	\begin{subfigure}{.45\textwidth}
		\centering
		\includegraphics[width=\linewidth]{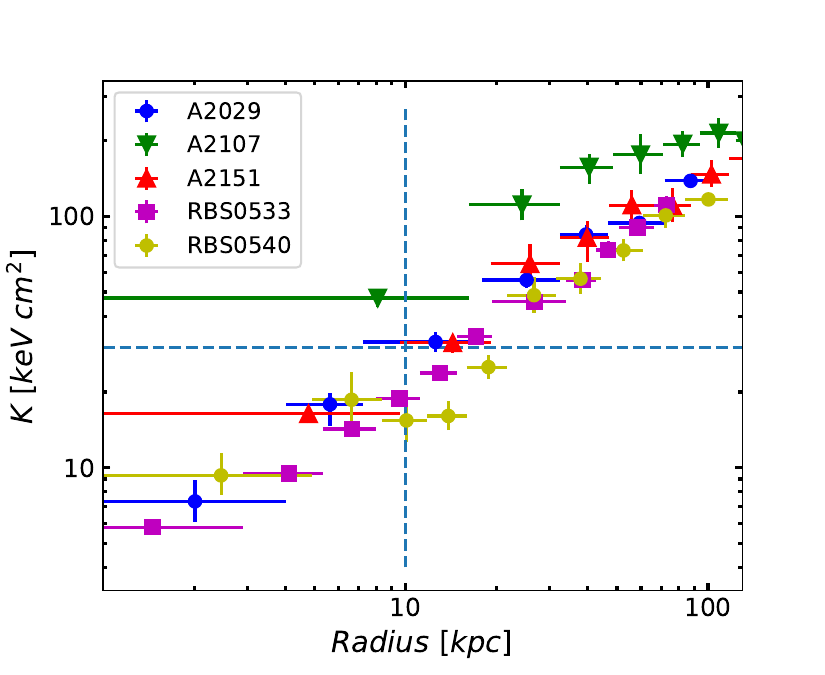}
		\caption{}
	\end{subfigure}
	\begin{subfigure}{.45\textwidth}
		\centering
		\includegraphics[width=\linewidth]{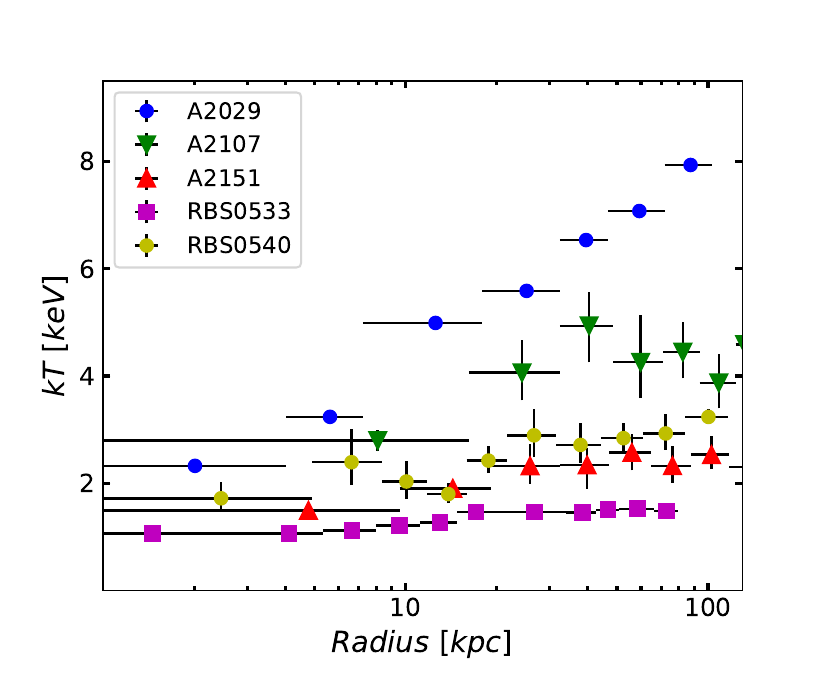}
		\caption{}
		\label{fig:kt_profile}
	\end{subfigure}
	\begin{subfigure}{.45\textwidth}
		\centering
		\includegraphics[width=\linewidth]{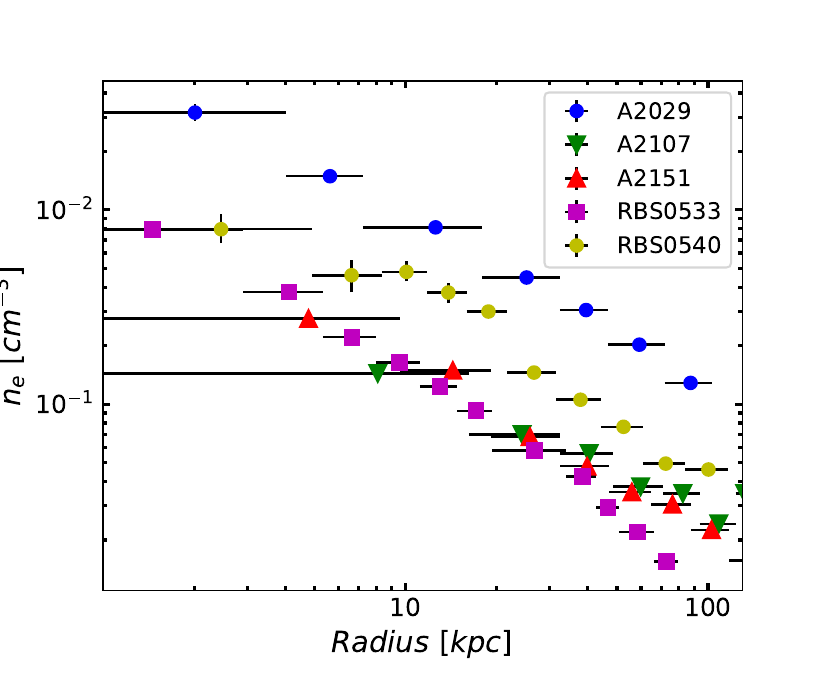}
		\caption{}
	\end{subfigure}
	\begin{subfigure}{.45\textwidth}
		\centering
		\includegraphics[width=\linewidth]{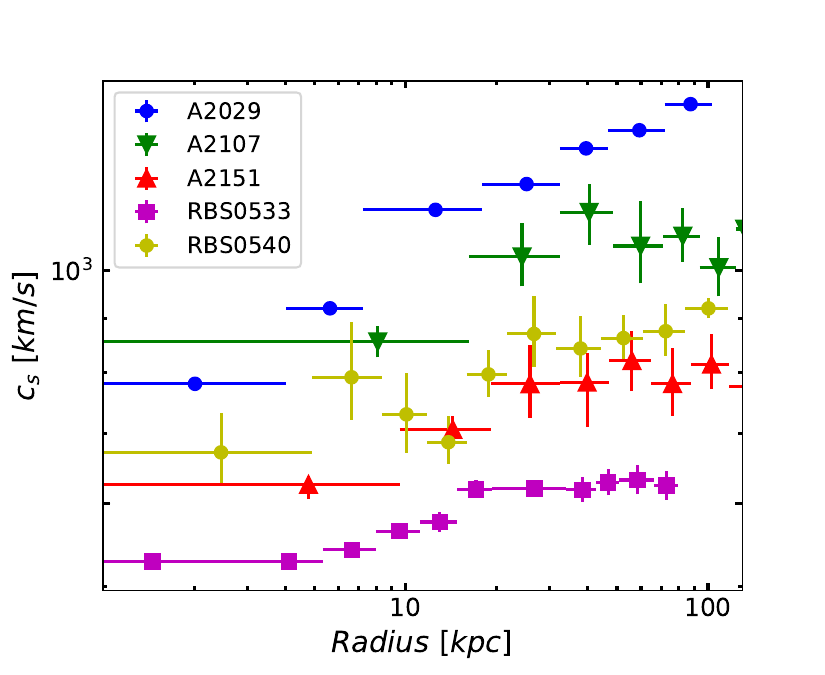}
		\caption{}
	\end{subfigure}

	\caption{Deprojected thermodynamic radial profiles in the $0.5 - 7$ keV band of sample clusters: (a) cooling timescale, (b) entropy, (c) electron temperature, (d) electron density, and (e) sound speed. The dashed vertical lines in (a) and (b) indicate a radius of 10 kpc, and the dashed horizontal lines represent the threshold values $\rm t_{cool} = 1.0 \times 10^9~yr$  and entropy $K = 30\ \rm keV\ cm^2$, respectively.}
	\label{fig:deprojected_profiles}
\end{figure*}

The deprojected radial profiles of (a) cooling timescale, (b) gas entropy, (c) temperature, (d) electron density, and (e) sound speed are depicted in Figure~\ref{fig:deprojected_profiles}. All five clusters reveal both short central cooling times, less than $\rm 10^{9}~yr$, and entropy lying below 30 $\rm keV~cm^2$ within the innermost 10 kpc, meeting the critical thresholds for bright nebular emission and high star formation rates. However, this does not align with our observations of smooth atmospheres, suggesting that additional processes are counteracting the expected cooling.

\subsubsection{Underlying Surface Brightness Model}
\label{sec: sb_model}

The surface brightness distributions are decomposed into two components: an `unperturbed' component, which typically peaks at the center and decreases with radius in relaxed clusters, and a `perturbed' component. The `unperturbed' SB distribution reflects the global potential of the cluster in equilibrium, and is commonly described by a spherically symmetric $\beta-$model:

\begin{equation}
    I_X(R)=I_0\left[1+\left(\frac{R}{r_c}\right)^2\right]^{-3\beta+0.5},
    \label{eq:sb_model}
\end{equation}
where $I_0$ is the central surface brightness, $r_c$ is the core radius, $R$ is the projected distance from the cluster center, and $\beta$ is a dimensionless parameter that determines the slope of the SB radial profile. The SB global gradient will dominate the power spectrum of raw X-ray images, resulting in high power at large scales and leakage to smaller scales. To analyze gas perturbations relative to this smooth `unperturbed' background, 
the raw images were divided by the best-fitting $\beta-$models to remove the large-scale SB gradient.

A spherical $\beta$ model does not capture the elliptical morphology of Abell 2029 or the slight west-east asymmetry observed in the SB of RBS0533. In addition, cool-core clusters have cusps of X-ray emission requiring a two-component $\beta$ model: one for the central peak and another for the broad, shallow outer gas distribution. To assess the uncertainty caused by using an oversimplified model, we re-analyzed the SB fluctuations in the Abell 2029 cool core with an elliptical double-$\beta$ model.

\begin{figure}
\centering
	\includegraphics[width=0.93\columnwidth,height=6.5cm]{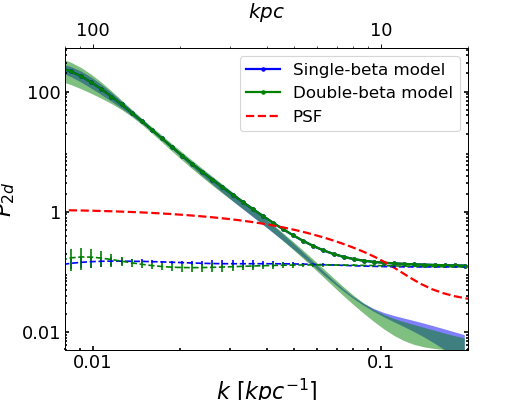}
    \caption{Power spectra of surface brightness fluctuations in the cool core of Abell 2029 derived using two different $\beta$-models. The global SB gradient has been removed by dividing the raw image with an elliptical single-$\beta$ model, marked in blue points, and a double-$\beta$ model, presented in green points. The horizontal dashed lines with error bars represent the Poisson noise levels, which were subtracted from the raw spectra to obtain the corrected power spectra of SB fluctuations, depicted by the shaded areas. The mean PSF spectrum, calculated from 20 simulated PSF images within the cool core of Abell 2029, is shown as a red dashed line.}
    \label{fig:2model_comparison}
\end{figure}

The comparison between two-dimensional (2D) power spectra of surface brightness fluctuations in the cool core of Abell 2029, obtained from the single $\beta-$model (in blue) and the double $\beta-$model (in green), is illustrated in Figure~\ref{fig:2model_comparison}. This comparison reveals that the power of SB fluctuations remains almost unchanged regardless of the underlying SB model used. However, a suppressed power amplitude is observed on large spatial scales when using the double-$\beta$ model. This suppression is attributed to the double $\beta$-model accounting more for features at large scales related to the underlying cluster potential rather than to perturbations. Small residuals between extracted SB profiles and best-fit profiles for both models (less than 3\%) indicate an acceptable fit of the single $\beta$-model for Abell 2029. The impact of the model choice on the SB analysis of Abell 2029 is negligible. However, the choice of model depends on the specific characteristics and the scales at which the analysis is performed and is therefore evaluated for each sample cluster.

A double $\beta$-model is used for analyzing 
Abell 2107 and RBS0540, while the single $\beta$-model is chosen for the others. The best-fit parameters are provided in Table~\ref{tab:beta_fitting} for each sample. As the innermost region surrounding the AGN was removed prior to fitting, the parameters listed in Table~\ref{tab:beta_fitting} do not describe the properties of the cool core and the periphery.

The uncertainties associated with selecting the underlying $\beta-$models are also explored in \citet{zhuravleva2015gas}. Their study concludes that while more complex models might suppress the amplitude of density fluctuations across a wider range, the measurements on relatively small scales remain largely consistent, except in cases where the gas is highly disturbed. This finding is consistent with ours. 

\begin{table*}
\centering
\begin{tabular}{lcccccccc}
\hline  
Cluster & $I_1$ & $R_{c,1}$ & $\beta_1$ & $I_2$ & $R_{c,2}$ & $\beta_2$ & $\psi$ & $e$  \\

        &   $\rm cnts/arcsec^{2}$  & arcsec  &   &  $\rm cnts/arcsec^{2}$  & arcsec &  & deg & \\

(a)& (b) & (c) & (d) & (e) & (f) & (g) & (h) & (i) \\        
\hline 
 RBS0533 & 50.2  & 8.2  & 0.40   & - & - & - & - & - \\
 Abell 2029   & 124.1 &  23.1 & 0.48   & - & - & - & 115 & 1.15  \\
 Abell 2151   & 5.5 & 11.7& 0.37 & - & - & - & -&- \\
 Abell 2107   & 7.3 & 44.3 & 6.2 & 3.2 & 36.7 & 0.37&- & -\\
 RBS0540 & 16.7 & 22.0 & 0.93 & 4.0 & 49.5 & 0.53 & -& -\\
\hline
\end{tabular}  
\caption{The best-fitting $\beta$-model parameters of the underlying surface brightness profiles of our targets: (a). Cluster name, (b-d). Central surface brightness, Core radius, and $\beta$ of the first component, (e-g). Central surface brightness, Core radius, and $\beta$ of the second component, (h). The orientation angle of the observed ellipse, and (i). Eccentricity, defined as the ratio of the major to the minor axis as projected.}
\label{tab:beta_fitting}
\end{table*}

\begin{figure*}
\centering
	\includegraphics[width=0.85\linewidth]
	{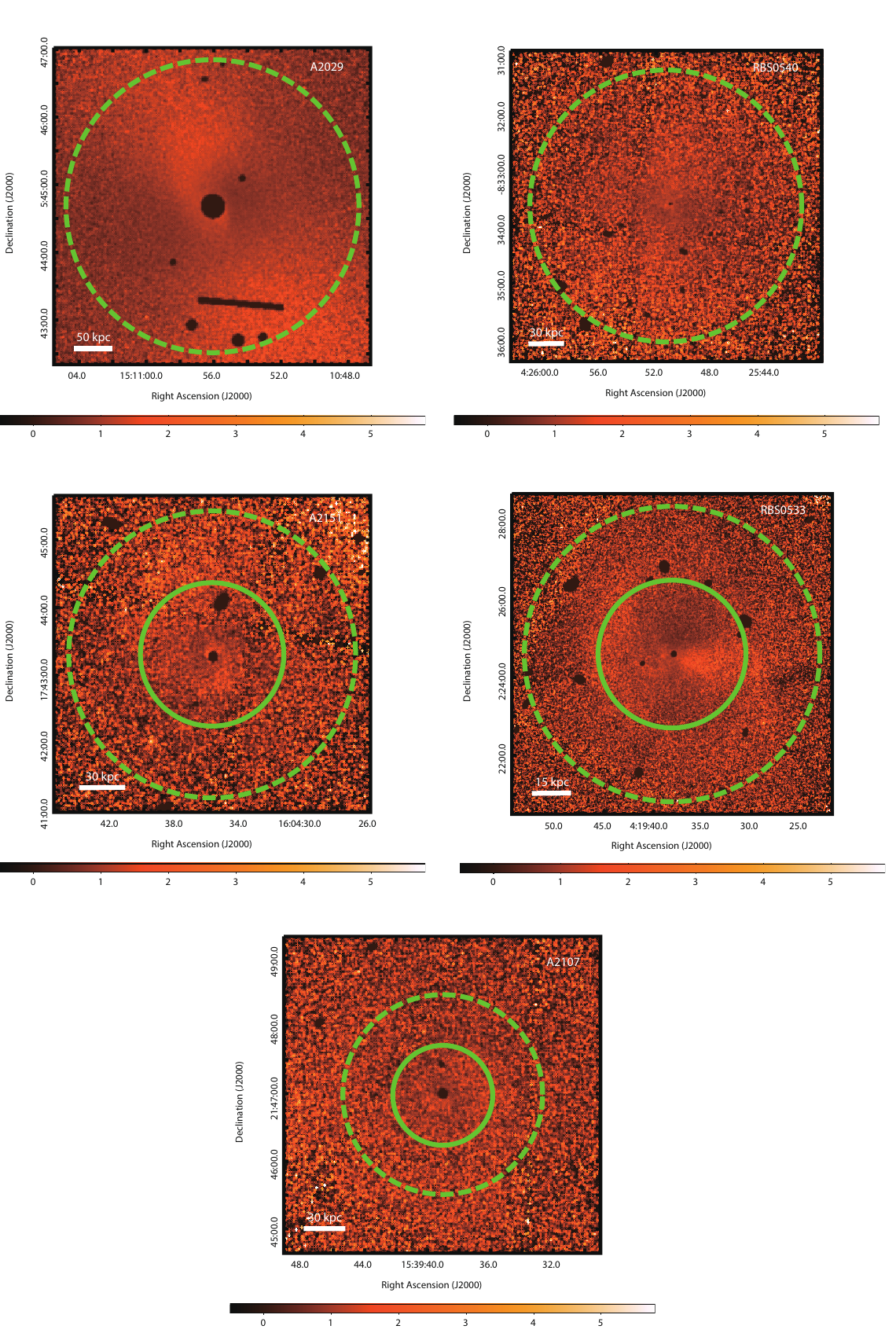}
    \caption{The residual images of smooth atmosphere clusters in the 0.5 - 3.5 keV band. The underlying surface brightness gradients are removed by dividing raw images by the best-fitting $\beta-$models, summarized in Table~\ref{tab:beta_fitting}. The cool-core radius is indicated with dashed green circles, and the inner half of the cool-core radius is marked in the green solid line. The residual images are processed with a 3" Gaussian smoothing for display purposes. The bright point sources and the central AGNs have been masked out.}
    \label{fig:residual_figure}
\end{figure*}

Figure~\ref{fig:residual_figure} shows the residual images of these smooth atmosphere clusters in the $0.5 - 3.5$ keV energy band, created by dividing the initial mosaic images shown in Figure~\ref{fig:chandra_figure_spoiler}, by the best-fitting spherically symmetric single/double $\beta-$model. Despite relatively structureless atmospheres, these X-ray residual images of sample clusters reveal structures that indicate disturbances in the ICM. RBS0533, for instance, shows a surface brightness depression to the north of its center, indicating a potential X-ray cavity. In contrast, the other four objects exhibit no prominent bubbles or cavities, typically associated with the central AGN activities. Notably, in Abell 2029, the residual image reveals a significant spiral-like feature extending clockwise from the center to about 400 kpc at its farthest extent \citep{paterno2013deep}. Such a substructure is likely a result of gas sloshing from cluster cores, induced by small infalling groups or subclusters that gravitationally perturb the cluster core, as evidenced by similar features in numerical simulations \citep{ascasibar2006origin,zuhone2010stirring,zuhone2011testing}. In Abell 2151, there appears to be a potential spiral feature, although it is not as distinct as the one observed in Abell 2029.

%% file: Cpt3_Power_spectral.tex

\section{Surface Brightness Fluctuation Analysis}
\label{sec: power_spectrum}

Gas perturbations in the X-ray residual images are quantified using the power spectrum of surface brightness fluctuations across various spatial scales. The $\Delta-$variance method as described by \citet{arevalo2012mexican} is used in this work, which is effective for images with irregular boundaries or significant areas missing. For a given spatial scale $k=1/\sigma\sqrt{2\pi^2}$, the images are convolved with two Mexican hat filters of widths $\sigma_1=\sigma/\sqrt{1+\epsilon}$ and $\sigma_2=\sigma\sqrt{1+\epsilon}$, with $\epsilon\ll 1$. This process isolates fluctuations at scale $\sigma$ by computing the differences between the two convolved images. The variances of the resulting images are then calculated as a function of the wavenumber, which serves as a measure of the power of the SB fluctuations $\rm P_{2D}(k)$ at a given spatial scale.

Poisson noise is expected to display a flat, white noise spectrum, especially dominating at smaller spatial scales. The power spectrum of Poisson noise, given the number of counts $\rm {n_{cts}}$ in each pixel, can be estimated, albeit with minor deviations due to the effects of image edges or data gaps. For this purpose, 100 mock images are generated, and each with Poisson noise simulated by multiplying the square root of the counts $\sqrt{\rm {n_{cts}}}$ in each pixel by a random number from a Gaussian distribution with a mean of 0 and a variance of 1. The PS of Poisson noise is calculated for each mock image and the average is used to estimate the Poisson noise contribution. As shown by the horizontal dashed lines in Figure~\ref{fig:2model_comparison}, the Poisson noise levels are comparable to the raw PS of SB fluctuations at large wavenumber k, thereby limiting our measurements at small spatial scales. We correct the Poisson noise by subtracting the estimated PS of Poisson noise from the raw PS. The scatter of these white noise spectra, denoted as $\rm \sigma_{{P_{wn}}}$, is then scaled by the following factor:

\begin{equation}
    {\rm \sigma_{{p}}={\sigma_{{p_{wn}}}}~\frac{{P_{2D}}}{{P_{wn}}}},
    \label{eq:scatter}
\end{equation}
giving an estimation of the uncertainties in PS of SB fluctuations. In this work, we present only the measurements at scales least affected, where the power of the cleaned spectra must exceed the level of Poisson noise.

In addition, the images are smeared by the instrumental PSF. Consequently, understanding the PSF at various positions across the cluster images is critical for correcting the suppressed power at small scales and accurately determining the high-k cutoff. We use a simplified analytic model to approximate the observed PSF spectra of the Lockman Hole field for the Chandra telescope, as provided by \citet{churazov2012x}:

\begin{equation}
    {\rm P_{PSF, Chandra}=\frac{1}{{\left[1+\left(\frac{k}{0.06}\right)^2\right]}^{1.1}}},
    \label{eq:psf_ps}
\end{equation}
where the unit of wavenumber k is $\rm arcsec^{-1}$. The PSF correction is applied by dividing the power spectra of the images by the power spectra of the PSF.

\begin{figure}
\centering
	\includegraphics[width=7cm,height=6.5cm]{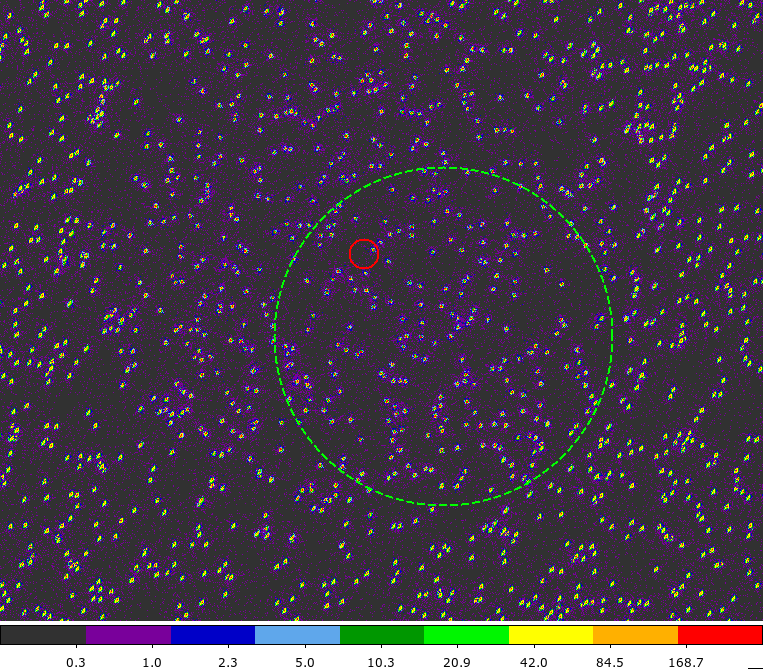}
    \caption{The simulated map of the combined Chandra PSF within the mosaic image of Abell 2029. The red solid circle indicates the position of the optical axis, while the dashed green circle represents the region of interest in our analysis.}
    \label{fig:psf_a2029}
\end{figure}

To accurately account for the PSF effects on the power spectrum, we generated simulated PSF maps for each cluster using the Chandra Ray Tracer (\texttt{ChaRT}) and \texttt{MARX} at various off-axis angles. The PSF for each observation is determined using Chandra's PSF libraries. The PSF images were then sampled across the field of view. The simulated map of the combined Chandra PSF of Abell 2029 is shown in Figure~\ref{fig:psf_a2029}. The PSF exhibits distortion, especially near the edges of the image. The power spectra of simulated PSF maps are computed and used for PSF effect correction. The mean PSF spectrum is calculated from 20 simulated PSF images within the cool core of Abell 2029, and is shown as a red dashed line in Figure~\ref{fig:2model_comparison}. 

The contribution of unresolved point sources on the power of SB fluctuations is estimated for each target by measuring the flux distribution of point sources, following the method given in \citep{churazov2012x,heinrich2024merger}. The power of unresolved point sources is much lower than that of bright point sources, which have been excluded from the residual images. In the cool core of Abell 2107, the ratio of the power of bright point sources $\rm P_{bright}$ to that of faint point sources $\rm P_{unresolved}$ is around $10^9$, while the ratio of other sample clusters is in range of $10^6 \sim 10^9$. Therefore, the contribution from the unresolved point sources is negligible to our analysis.

The X-ray emissivity is proportional to the integral of the square of the electron number density ($\rm \int n_e^2 dl$) over the line of sight. Therefore, in such clusters, the amplitudes of SB fluctuations indicate the amplitudes of gas density fluctuations. We converted the 2D SB fluctuations power spectra, $\rm P_{2D}(k)$, into 3D density fluctuations power spectra, $\rm P_{3D}(k)$, assuming a spherically symmetric geometry of clusters. This process is discussed in detail in the works of \citet{churazov2012x} and \citet{zhuravleva2015gas}. The conversion factor between the 3D power spectrum and the 2D power spectrum at each line of sight $z$ is given by the following relationship:

\begin{equation}
    {\frac{P_{2D}(k)}{P_{3D}(k)}\approx 4\int \left|W(k_z)^2\right|dk_z},
    \label{eq:conversion}
\end{equation}
where $\left|W(k_z)\right|^2$ is the 1D power spectrum of the normalized emissivity distribution along the line of sight. The dependence of this normalization constant $\int \left|W(k_z)^2\right|dk_z$ in the x and y direction can be neglected in the central regions of clusters with `flat' SB profiles, such as the Coma cluster. In our sample cool-core clusters, the SB slope is usually steep in the center and this correcting factor will vary across the region of interest. Therefore, in this work, we implemented a `local' correction factor to account for these variations.

Rather than using the gas density power spectra $\rm P_{3D}(k)$ directly, it is often more convenient to employ the characteristic amplitude of density fluctuations:

\begin{equation}
\frac{\delta \rho_k}{\rho_0} \equiv A_{3D}(k) = \sqrt{4\pi P_{3D}(k) k^3},
\label{eq:a3d}
\end{equation}
which represents the typical variations in gas density relative to the `unperturbed' smooth density distribution, since its unit is the same as that of the variable in real space.

The velocity power spectra for gas motion can be inferred from the power spectra of density fluctuations. In a stratified ICM where perturbations are small, the one-component velocity at each wavenumber $k$, denoted by $V_{1,k}$, is linearly proportional to the amplitude of density fluctuations $\rm A_{3D}(k)$:

\begin{equation}
\frac{\delta \rho_k}{\rho_0}=\eta_1\frac{V_{1,k}}{c_s},
\label{eq:a3d_vturb_relation}
\end{equation}
where $\rm c_s$ represents the sound speed in the atmosphere as given in equation~(\ref{eq:sound_speed}). \cite{zhuravleva2014relation} provided a theoretical argument and confirmed this linear scaling relation across a broad range of scales in both buoyancy-dominated and turbulence regimes using cosmological simulations of relaxed clusters. The proportionality coefficient $\eta_1 = 1 \pm 0.3$ at scales of 30 - 300 kpc. Using simulations of 80 clusters in different dynamical states, \cite{zhuravleva2023indirect} verified this strong linear correlation between the gas density fluctuation amplitudes and the Mach number of gas motions in relaxed galaxy clusters. After accounting for the ellipticity of the gas distribution, the average $\eta_1$ is $ 0.9\pm 0.2~(1.2\pm 0.3)$ at scales from 60 to 300 kpc. Our clusters are well-represented by the simulated relaxed clusters in \cite{zhuravleva2023indirect}. While simulations validate the linear scaling relation across various conditions, they typically do not consider factors such as metallicity variations and non-thermal, spatially variable components. These elements can induce emissivity fluctuations independent of gas motion velocities, implying that the derived scale-dependent velocity amplitudes might represent upper limits.

%% file: Cpt4_Results.tex

\section{Results}
\label{sec:results}
Here we present our measurements of gas density fluctuations in Section~\ref{sec: density_fluctuations}. We discuss indirect measurements of one-component velocity spectra derived from the power spectra of density fluctuations in Section~\ref{sec: velocity_ps}. In Section~\ref{sec: turbulent_heating}, the turbulent heating rate for each object is estimated and compared to the radiative cooling rate to examine if turbulent dissipation is able to balance cooling in the cluster cores.

\subsection{Density Fluctuations in the Smooth Atmosphere Clusters}
\label{sec: density_fluctuations}

\begin{figure*}
	\begin{subfigure}{.32\textwidth}
		\centering
		\includegraphics[width=\linewidth]{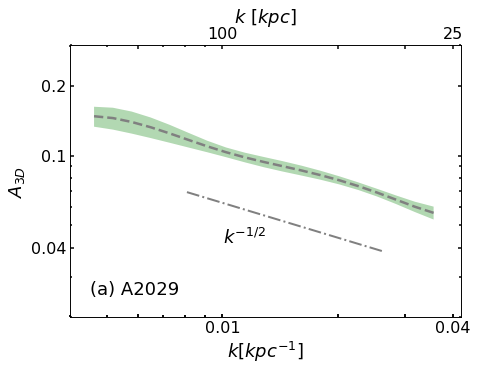}
		\label{fig:A3D_A2029}
	\end{subfigure}
 	\begin{subfigure}{.32\textwidth}
		\centering
		\includegraphics[width=\linewidth]{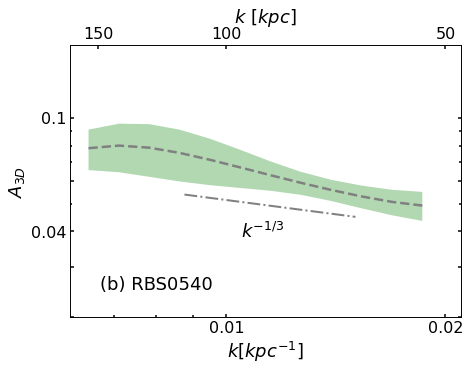}
		\label{fig:A3D_RBS0540}
	\end{subfigure}
 	\begin{subfigure}{.32\textwidth}
		\centering
		\includegraphics[width=\linewidth]{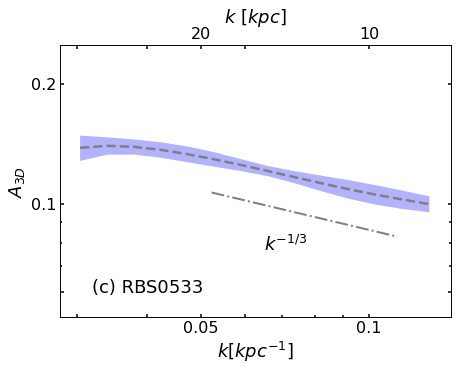}
		\label{fig:A3D_RBS0533}
	\end{subfigure}
	\begin{subfigure}{.32\textwidth}
		\centering
		\includegraphics[width=\linewidth]{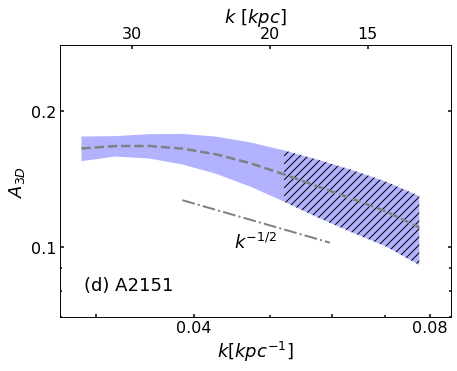}
		\label{fig:A3D_A2151}
	\end{subfigure}
 	\begin{subfigure}{.32\textwidth}
		\centering	
        \includegraphics[width=\linewidth]{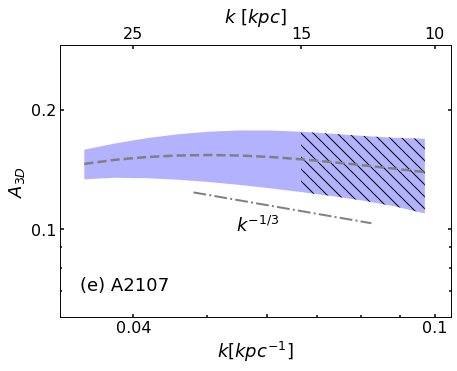}
		\label{fig:A3D_A2107}
	\end{subfigure}
	\begin{subfigure}{.32\textwidth}
		\centering
		\includegraphics[width=\linewidth]{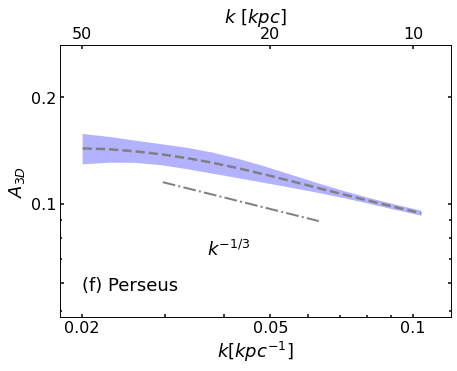}
		\label{fig:A3D_Perseus}
	\end{subfigure}
	\caption{Gas density Fluctuation amplitudes of our five sample clusters and the Perseus Cluster. The measurements were conducted within the cool-core radius of (a) Abell 2029, (b) RBS0540, and within the inner half of the cool-core radius of (c) RBS0533, (d) Abell 2151, (e) Abell 2107, and (f) the Perseus Cluster. The central dashed lines present the mean spectra, while the widths of the shaded regions reflect the $1\sigma$ uncertainties. Displayed are measurements taken over scales that are minimally impacted by Poisson noise and PSF effect < 20\%. The results below the scale of 15 kpc for Abell 2107 and 20 kpc for Abell 2151, marked with hatched areas, maybe more significantly affected by PSF effects.The grey dash-dotted lines represent slopes of $k^{-1/3}$ or $k^{-1/2}$ as guidelines.}
	\label{fig:ps_density_fluctuations}
\end{figure*}

The power spectra of surface brightness fluctuations in the $0.5 - 3.5$ keV band are measured to determine the density fluctuation amplitudes as a function of the wavenumber $k$. Figure~\ref{fig:ps_density_fluctuations} presents the power spectra of density fluctuations measured in the cool cores of Abell 2029, RBS0540, and within the inner half cool-core radius of RBS0533, Abell 2151, Abell 2107, and Perseus. The mean spectra are shown as the dashed lines, while the widths of the hatched regions represent the $1\sigma$ statistical uncertainties.

We excluded the measurements at large scales where amplitude flattening occurs. This flattening likely indicates leakage from the global surface brightness gradient induced by uncertainties in selecting the appropriate SB model, discussed in Section~\ref{sec: sb_model}. The smallest scales that can be probed are determined by both the Poisson noise level and PSF distortions. The raw spectra of SB fluctuations are corrected using the estimated Poisson noise level. If the cleaned power spectrum lies below the power of Poisson noise, those measurements are excluded, and only the measurements over scales least affected by noise are retained. In addition, while equation~(\ref{eq:psf_ps}) implies that PSF blurring effects are negligible with Chandra instruments at the low-k end ($P_{\rm PSF, Chandra} \sim 1$, presented in Figure~\ref{fig:2model_comparison}), our analysis of the PSF effect, conducted using simulated PSF maps for each cluster as detailed in Section~\ref{sec: power_spectrum} reveals variations. After PSF correction, we only accept measurements over scales where the PSF suppression on 2D power spectra is less than 30\%, which corresponds to less than 20\% suppression on density fluctuation amplitudes. This condition ensures the reliability of the results, but measurements over scales less than 10 kpc cannot be achieved in the core regions of all targets, with one exception, RBS0533, where the measured gas density fluctuations at the scale of 8 kpc is 10 per cent. The high-wavenumber cutoff in its inner half cool-core is determined by the Poisson noise level, even when using strict criteria. Additionally, the PSF suppression is $\sim$ 20\% at 10 kpc due to its low redshift of 0.0123 (lower than Perseus z = 0.0176). Furthermore, its cooling region is small, with $\rm r_{cool} \sim 60$ kpc, and we only analyzed the inner half core. This means the region of interest is the very central part, where the PSF effect is relatively small. To maximize the utilization of the data, we have retained the results below the scale of 15 kpc for Abell 2107 and 20 kpc for Abell 2151 in Figure~\ref{fig:ps_density_fluctuations}, marked with hatched areas, even though the measurements may be more significantly affected by PSF effects.

\renewcommand{\arraystretch}{1.3} 
\begin{table}
\centering

\begin{tabular}{p{4.5cm}p{4.5cm}p{4.5cm}p{5cm}}
\hline
Cluster & Region   & Scales        & $\frac{\delta\rho}{\rho_0}$             \\
        & kpc      & kpc           & per                                     \\
(a)     & (b)      & (c)           & (d)                                        \\ \hline
RBS0533 & 1.5 - 30 & 8 $\sim$ 33   & 10.0 $\sim$ 13.8          \\
Abell 2029   & 5 - 180  & 28 $\sim$ 216 & 5.7 $\sim$ 14.8          \\
Abell 2151   & 3 - 47   & 20 $\sim$ 35  & 14.5 $\sim$ 16.5           \\
Abell 2107   & 4 - 35   & 15 $\sim$ 29   & 14.6 $\sim$ 15.3         \\
RBS0540 & 5 - 115  & 54 $\sim$ 155 & 4.9 $\sim$ 7.8        \\
Perseus & 5 - 87   & 12 $\sim$ 50   & 10.2 $\sim$ 14.4    \\ \hline
\end{tabular}
\caption{The density fluctuation amplitudes measured within the cooling regions of our sample clusters and within the inner half cool core region of Perseus: (a). Cluster name, (b). Region of interest, (c). Spatial scales least affected by Poisson noise,  (d). Amplitudes of gas density fluctuations.}
\label{tab:density_fluctuations}
\end{table}


Although the ranges of measurable scales vary across the objects, the typical amplitudes of density fluctuations remain relatively small. In four out of the five clusters, the amplitudes are less than 15 per cent even on the largest scales probed. These modest amplitudes suggest mild gas perturbations. The measured gas density amplitudes are summarized in Table~\ref{tab:density_fluctuations}. An exception is observed in Abell 2151, where it reaches $\sim 17$ per cent at a scale of 35 kpc. 

The 1.4 Ms of Chandra observations of the Perseus cluster enable us to probe surface brightness fluctuations at scales of $\sim 12$ kpc. Within the inner half cool-core region, the SB amplitudes vary from $\sim$ 10 per cent to 14 per cent at scales of 12 - 50 kpc. The residual image of the Perseus cluster in the right panel of Figure~\ref{fig:chandra_figure_perseus} reveals that a significant portion of its core is filled with bubbles and sharp shock edges. These features are driving the relatively high fluctuation amplitudes. Similarly, the residual image of RBS0533 displays a surface brightness depression north of its center, identified as a possible X-ray cavity of $\sim$ 7.5 kpc in size, $\sim$ 10 kpc from the center by \citet{martz2020thermally}. The bubble in RBS0533 is potentially capable of inducing gas turbulent motions comparable to those observed in Perseus, with amplitudes ranging from $\sim$ 9 to 14 per cent at scales of 5 - 33 kpc.

Despite the overall lack of X-ray cavities in structureless atmospheres, the observed density fluctuations are comparable to those in the Perseus cluster in four of five of our systems, except RBS0540. This suggests that other factors, such as gas sloshing, create gas perturbations. In Abell 2029, the amplitudes of density fluctuations are $\sim 5.7-14.8$ per cent at scales of $28 - 216$ kpc. In contrast, Abell 2151 exhibits higher density fluctuations in its core, ranging between $14.5 - 16.5$ per cent at scales of $20 - 35$ kpc, surpassing those observed in Perseus, which is known for its feature-rich X-ray images. The residual image of Abell 2029 clearly reveals a spiral feature related to a cold front, as indicated by a sharp change in surface brightness. A less distinct sloshing feature observed in the residual image of Abell 2151 suggests the possibility of an ongoing merger. These observations support the hypothesis that the gravitational disturbances induced by merging subclusters may also effectively trigger gas density fluctuations in these systems.

Due to the relatively high redshift and limited exposure, density fluctuations in our targets cannot be measured at the small scales that are possible in Perseus, complicating direct comparisons. Nonetheless, in RBS0540 the fluctuation amplitudes are lower than in the other four objects, remaining below 10 per cent across the probed range of spatial scales. These amplitudes could be even lower at smaller scales or within the outer regions where the gas is less disturbed than the core regions.

\subsection{Velocity Power Spectrum}
\label{sec: velocity_ps}

\begin{table*}
\begin{tabular}{lccccccc}
\hline
Cluster & $M_{2500}$ & Region & T          & Scales        & $\rm V_{1,k}$     & $\rm Q_{turb}$             & $\rm Q_{cool}$             \\
        & $10^{14}~M_{\odot}$ & kpc & keV   & kpc           & $\rm km~s^{-1}$           & $\rm erg\ cm^{-3}\ s^{-1}$ & $\rm erg\ cm^{-3}\ s^{-1}$ \\
(a)     & (b) & (c)       & (d)           & (e)            & (f)                    & (g)             & (h)       \\ \hline
RBS0533 & $0.17^{+0.02}_{-0.02}$ & 2 - 30 & 1.2   & 8 $-$ 33   & 60 $-$ 83   & $9.9\times10^{-28}$    & $9.2\times10^{-28}$    \\ [8pt]
Abell 2029   & $5.1^{+0.20}_{-0.18}$ & 4 - 90 & 6.9    & 16 $-$ 105 & 68$-$ 162   & $2.1\times10^{-27}$    & $9.5\times10^{-27}$    \\[8pt]

         & &90 - 180 & 8.4   & 44 $-$ 217 & 92$-$ 147   & $3.2\times10^{-28}$    & $9.6\times10^{-28}$    \\[8pt]

Abell 2151   & $5.1^{+0.20}_{-0.18}$ & 4 - 47 &2.2   & 20 $-$ 35  & 112 $-$ 126  & $1.8\times10^{-27}$    & $1.1\times10^{-27}$    \\[8pt]
Abell 2107   & $1.05^{+0.23}_{-0.23}$ & 3 - 35 &3.9  & 15 $-$ 29   & 169 $-$ 178 & $4.9\times10^{-27}$    & $1.4\times10^{-27}$    \\[8pt]
RBS0540 & $0.32^{+0.03}_{-0.03}$ & 4 - 55 &2.8  & 54 $-$ 155 & 43 $-$ 68   & $6.1\times10^{-27}$    & $6.5\times10^{-29}$    \\[10pt]
Perseus & $1.5^{+0.03}_{-0.03}$ & 6 - 90 &3.8   & 12 $-$ 50   & 76 $-$ 143  & $1.1\times10^{-26}$    & $1.4\times10^{-26}$    \\\hline
\end{tabular}
\caption{One-component velocity amplitudes of gas motion measured indirectly in the cooling regions of our targets and Perseus using SB fluctuation analysis: (a). Cluster name, (b). Cluster mass $M_{2500}$, (c). The region of interest, (d). Volume-weighted mean temperature in the region of interest, (e). Scales probed, (f). One-component velocity, (g). Turbulent heating rate is calculated when volume-weighted mean temperatures and densities are used, (h). Radiative cooling rate.}
\label{tab:velocity_spectra}
\end{table*}

\begin{figure*}
	\begin{subfigure}{.32\textwidth}
		\centering
		\includegraphics[width=\linewidth]{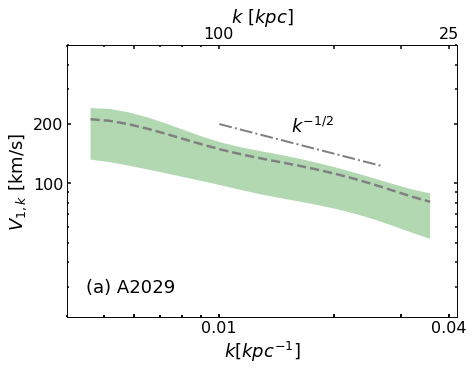}
		\label{fig:Vk_A2029}
    \end{subfigure}
  	\begin{subfigure}{.32\textwidth}
		\centering
		\includegraphics[width=\linewidth]{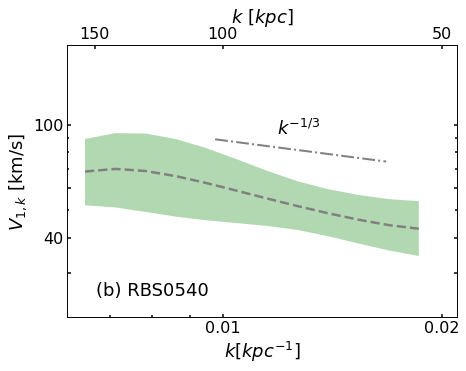}
		\label{fig:Vk_RBS0540}
	\end{subfigure}
	\begin{subfigure}{.32\textwidth}
		\centering
		\includegraphics[width=\linewidth]{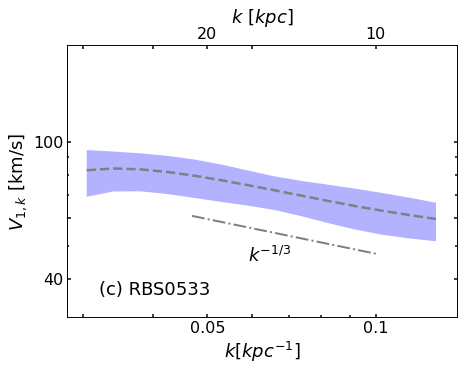}
		\label{fig:VK_RBS0533}
	\end{subfigure}
 	\begin{subfigure}{.32\textwidth}
		\centering
		\includegraphics[width=\linewidth]{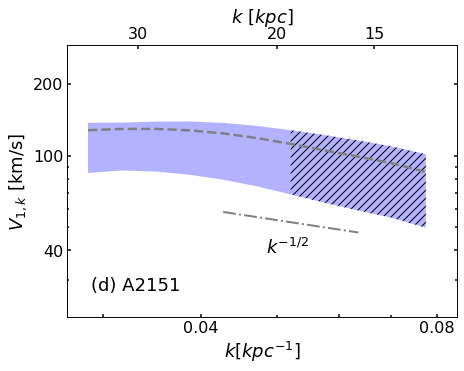}
		\label{fig:Vk_A2151}
	\end{subfigure}
	\begin{subfigure}{.32\textwidth}
		\centering
		\includegraphics[width=\linewidth]{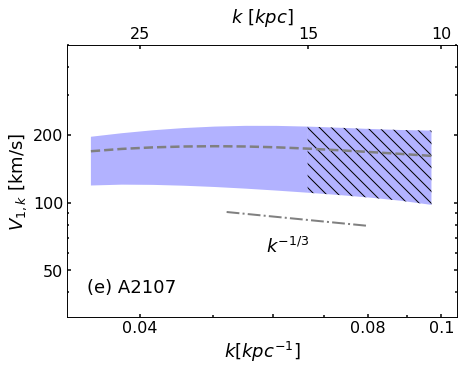}
		\label{fig:Vk_A2107}
	\end{subfigure}
	\begin{subfigure}{.32\textwidth}
		\centering
		\includegraphics[width=\linewidth]{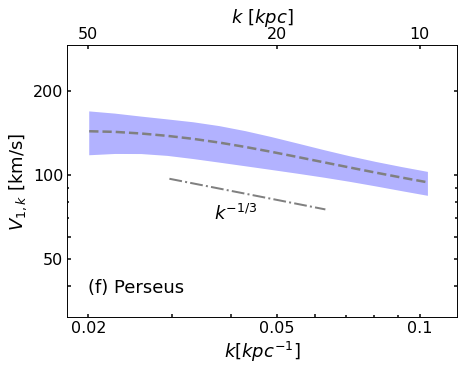}

		\label{fig:Vk_Perseus}
	\end{subfigure}
	\caption{Amplitudes of one-component velocity with respect to wavenumber k within the cool-cores in (a) Abell 2029, (b) RBS0540, and within the inner half cool-core regions of (c) RBS0533, (d) Abell 2151, (e) Abell 2107, and (f) Perseus cluster. Velocities derived from the average sound speed in the ICM are shown as the central dashed lines. The velocity estimation below the scale of 15 kpc in Abell 2107 and 20 kpc in Abell 2151, marked with hatched areas, might be more significantly affected by PSF effects. The grey dash-dotted lines represent slopes of $k^{-1/3}$ or $k^{-1/2}$ for guidance.}
	\label{fig:velocity_ps}
\end{figure*}

Using the measured power spectra of density fluctuations, the amplitudes of the one-component velocity of gas motions can be derived indirectly, following the linear scaling relation illustrated in equation~\ref{eq:a3d_vturb_relation}. The one-component velocity spectra in the cooling regions of our targets and Perseus are presented in Figure~\ref{fig:velocity_ps}. The dashed lines show the mean velocity spectra, obtained by using the volume-weighted sound speed of the gas, while the widths of the regions represent the $1\sigma$ uncertainties, which reflect the variances in sound speed across the cool cores of Abell 2029, RBS0540, as well as within the inner half cool-cores in RBS0533, Abell 2151, Abell 2107, and Perseus. The temperatures used to calculate the mean sound speed and the measured velocities are summarized in columns (d) and (e) of Table~\ref{tab:velocity_spectra}, respectively.

In the inner half-cool-core of Perseus, the velocities of gas motion range from $76~ \rm km ~s^{-1}$ at 12 kpc to $143~ \rm km ~s^{-1}$ at 50 kpc, with some scatter. Our indirect velocity measurements are also compared with direct measurements from the Hitomi Satellite in Perseus's central 30 - 60 kpc region, where velocities varied between $110~ \rm km ~s^{-1}$ at the scale of 5 kpc to $202~ \rm km ~s^{-1}$ at the scale of 30 kpc. Notably, both indirect and direct measurement methods yield consistent results, despite the uncertainty of the dominant scale measured by Hitomi. A similar comparison in Perseus was presented by \cite{zhuravleva2018gas}, aligning with our results. The upcoming XRISM observations, capable of directly measuring line-of-sight gas velocities, will provide further testing and refine the relationship between density perturbations and gas motion velocity.

Despite a potential cavity structure in RBS0533, which contributes to higher density fluctuations, the measured velocities are relatively low, ranging between 60 $\rm km ~s^{-1}$ and 83 $\rm km ~s^{-1}$ at the scale of $8 - 33$ kpc. This is likely due to its low central temperature, which is less than 2 keV. Similarly, the structureless cluster RBS0540 also exhibits velocities of below  100 $\rm km ~s^{-1}$. In contrast, the velocities measured in Abell 2029, Abell 2151, and Abell 2107 are comparable to those observed in Perseus, all exceeding 100 $\rm km ~s^{-1}$, aligning with the expected turbulent velocities in the ICM. Notably, Abell 2029, which exhibits a spiral structure, shows velocities between 81$~\rm km ~s^{-1}$ and $211~\rm km ~s^{-1}$ at scale of $28 - 217$ kpc, whereas Abell 2151 displays velocities of $112 - 126~\rm km ~s^{-1}$ at scale of $20 - 35$ kpc.

\subsection{Turbulent Heating}
\label{sec: turbulent_heating}

The turbulent heating rate is estimated from the inferred turbulent velocity calculated using the relation ${\rm Q_{turb} }\propto \rho_0 V^3_{1,k}/l$. Here, $\rho_0$ represents the mean gas density defined as $\rho_0=(n_e+n_i)\mu m_p$, where $n_e$ and $n_i$ are the electron and ion number densities, respectively. $\mu$ is the mean molecular weight, and $m_p$ is the mass of a proton. The scale-dependent velocity spectra $V_{1,k}$ are shown in Figure~\ref{fig:velocity_ps}, and $l$ is the scale within the inertia range on which the $V_{1,k} \propto l^{1/3},~(k\propto 1/l)$. The turbulent heating rate is then derived using:

\begin{equation}
    {\rm Q_{turb} =\rho_0~\epsilon=C_Q~\rho_0~V^3_k~k},
\end{equation}
where $\rm C_Q = 3^{3/2}~2\pi / (2C_K)^{3/2} \approx 5$ is a dimensionless constant where the Kolmogorov constant \citep[]{sreenivasan1995universality}, $\rm C_K$, is $\sim 1.65$. To assess the efficiency of turbulent heating in each cluster, we compare the turbulent heating rate with the gas cooling rate. The radiatively cooling rate then is calculated as $\rm Q_{cool} = n_e n_i \Lambda(T)$, where $\Lambda(T)$ is the normalized gas cooling function for 0.3 solar abundance.

\begin{figure*}
    \centering
    \includegraphics[width=17cm,height=9cm]{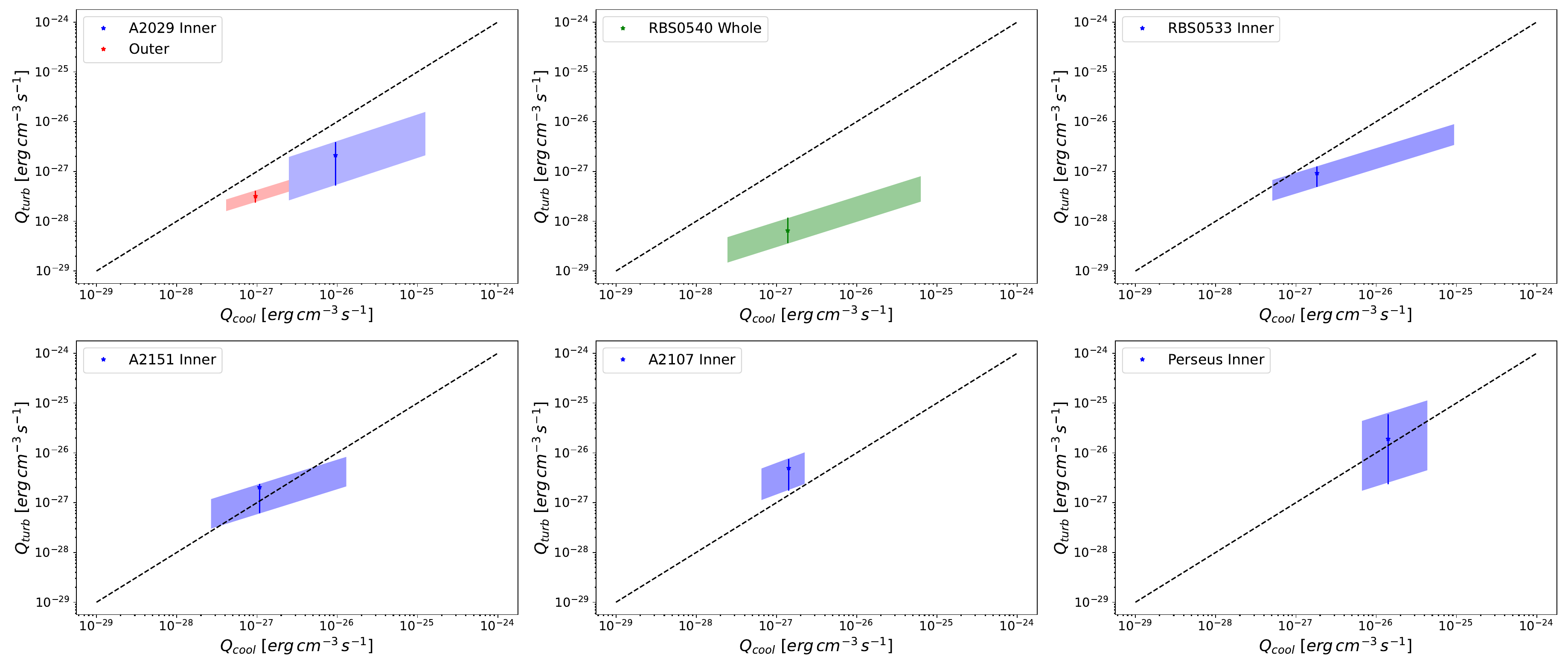}
    \caption{Turbulent heating versus radiative cooling rates within the cool-core radius of Abell 2029 and RBS0540, and within the inner half cool-cores of RBS0533, Abell 2151, Abell 2107, and Perseus. For Abell 2029, the comparison is presented separately for the inner (blue) and outer (red) cool-core regions. The central points represent the $\rm Q_{cool}$ and $\rm Q_{turb}$ obtained using volume-weighted temperatures and densities. The error bars indicate $1\sigma$ uncertainties in heating rate, variations in density and temperature within the regions of interest, and deviations from the Kolmogorov spectral slope of $k^{-1/3}$. The dashed line illustrates the equilibrium where radiative cooling is balanced by turbulent heating.}
    \label{fig: heating_cooling_rate}
    \end{figure*}

Figure~\ref{fig: heating_cooling_rate} presents a comparison between the heating and cooling rates within the cool cores of Abell 2029 and RBS0540 and within the half cool-core radius of Abell 2107, Abell 2151, RBS0533, and Perseus. The dashed lines denote equality between the heating rate and cooling rate. The central dots represent the heating and cooling rates calculated using volume-weighted mean gas temperature and density within the regions of interest. The size of each rectangle represents the $1\sigma$ uncertainties including variations in gas temperature/density across the regions and the statistical uncertainties in measured velocity spectra. Significant uncertainties in the cooling rates arise due to the variations in gas density and temperature, which can stretch the error range and mask the tension between heating and cooling rates. To address this, we have compared the inner and outer cool core regions of Abell 2029 separately (shown in blue and red, respectively). Both the heating and cooling rates are higher in the inner cool core compared to the outer cool core, with the larger uncertainty resulting from the steeper density profile in the cluster center.

Assuming Kolmogorov-like turbulence, the cascade rate $\rm Q_{turb} \sim V_{1,k}^3k$ should remain approximately constant across the inertial range with a velocity spectrum $V_{1,k} \propto k^{-1/3}$. However, the velocity spectra slopes of our sample clusters depart from the Kolmogorov slope of $-1/3$, enlarging in the uncertainties in $\rm Q_{turb}$. In addition, the uncertainty also arises due to the calibration of coefficient $\eta_1$ when we derived velocity spectra from the spectra of gas density fluctuations. In our analysis, we acknowledge that there is an order of magnitude uncertainty in the heating rate estimation.

Given its short central cooling time and low entropy, RBS0533 is expected to exhibit bright nebular emission. However, its $\rm H_\alpha$ luminosity lies below $\rm 0.016 \times 10^{40}\ erg\ s^{-1}$ in the ACCEPT database. \citet{o2018cold} found an upper limit to the molecular hydrogen mass at $\rm M_{H_2} < 0.47 \times 10^{8}\ M_\odot$. \citet{martz2020thermally} suggested that the bubble in RBS0533 can't lift enough cool gas to approximately 40 kpc from the cluster center where $\rm t_{cool}/t_{ff}<1$, thus failing to meet the cooling instability criterion and resulting in the absence of nebular emission. Moreover, as illustrated in Figure~\ref{fig: heating_cooling_rate}, the turbulence-induced energy within the inner half cool-core radius of RBS0533 has the potential to supply a substantial portion of the heat necessary to balance the radiative cooling. Therefore, dissipation turbulence injected by other sources might be heating the atmosphere. Our measurements are consistent with heating from the dissipation of turbulent gas motions being comparable to the radiative cooling in all systems apart from RBS0540.

%% file: Cpt5_Discussion.tex

\section{Discussion}
\label{sec: discussion}

\subsection{The Gas Perturbation Sources in Smooth Atmosphere Clusters}
\label{sec: driving_source}

The existing Chandra archival observations used in this work allow us to constrain density fluctuations over a limited range of spatial scales. Consequently, a direct comparison between these structureless systems and the Perseus cluster over identical scale ranges is infeasible. Nevertheless, based on the results shown in Figure~\ref{fig:ps_density_fluctuations}, it can be inferred that typical density fluctuations within the clusters with smooth atmospheres are generally modest across the range of measurable scales. Noteworthy exceptions are observed in Abell 2151 which range from $\sim$ 10 per cent to 17 per cent at scales of $12 - 35$ kpc, and in RBS0540, ranging from $\sim$ 6 per cent to 8 per cent at scales of $33 - 155$ kpc. By comparison, the core of the Perseus cluster exhibits density fluctuations ranging between 8 and 14 percent at scales of $4 - 50$ kpc, suggesting that the perturbation levels in Perseus are comparable to, or slightly exceed, those in tranquil-core clusters. The absence of pronounced bubble structures in our targets, perhaps apart from RBS0533, raises questions about the underlying mechanisms responsible for generating the observed levels of gas perturbations in these systems.

In the central region of Abell 2029, the ratio of cooling time to free-fall time consistently remains above 15 - 20, suggesting a thermally stable environment but one that is susceptible to precipitation. Nevertheless, a heating mechanism is required to account for the absence of cooling products such as nebular emission and star formation. Studies of turbulent dissipation in the central regions of more than ten nearby brightest cool-core clusters, including Perseus, have indicated that turbulence can sufficiently offset radiative cooling on average \citep{zhuravleva2018gas}. Central AGN activity likely plays a role by transferring energy to the ICM through turbulence induced by jets and bubbles.

Despite the absence of prominent X-ray cavities, we infer from Figure~\ref{fig: heating_cooling_rate} that the dissipation of turbulence in the core of Abell 2029 somewhat balances its cooling rate. The cD galaxy in Abell 2029, IC1101, harbors a relatively powerful radio source, PKS 1508+059 \citep{taylor1994searching,becker1995first}, with jet power of $P_{1.4} \sim 10^{41}\rm erg~s^{-1}$ and a size of $\sim$15 - 22 kpc. Radio sources of comparable power and size have been known to inflate X-ray cavities or bubbles. If PKS 1508+059 is indeed heating the ICM in Abell 2029 without creating bubbles, we would expect to observe stronger fluctuations near the radio source and lower amplitudes further away. We measured the inner and outer cooling regions separately in Abell 2029's core and observed higher gas density fluctuations in the inner regions. At a scale of $\sim$ 100 kpc, the density fluctuation is around 12 per cent in the inner region and 10 percent in the outer region. However, comparisons over smaller scales cannot be conducted. The reason why smooth atmospheres, such as Abell 2029, exhibit gas perturbation levels similar to those in Perseus remains unclear. To accurately measure density fluctuation amplitude down to the scale of the radio source, a substantially longer exposure of approximately 1.4 Ms for Abell 2029 would be necessary.

An alternative possibility is that gas sloshing induces density fluctuations. A prominent spiral is clearly visible in Abell 2029, identified as a cold front. The relaxed atmosphere of Abell 2029 shows no obvious indication of a major merger. The cold front is likely created by a `slosh' motion induced by merging halos, which is displacing low-entropy gas from the cluster cores. This sloshing is likely to disturb the ICM and heat the cluster core by mixing hot gas from the outer regions with the cooler gas in the core. \citet{zuhone2010stirring} used high-resolution N-body/Eulerian hydrodynamic simulations to model gas sloshing initiated by mergers with subclusters in galaxy clusters closely resembling Abell 2029. Their results indicate that gas sloshing can facilitate the inflow of heat to the cluster core by mixing hot gas from the cluster outskirts with the cool-core gas. This process redistributes gas in cooling cores, thereby reducing the efficiency of radiative cooling. The impact of gas sloshing varies with viscosity and the frequency of mergers, but they suggest that a cooling catastrophe can be prevented for intervals of approximately 1–3 Gyr.

 \citet{walker2018fraction} conducted an analysis of the Perseus cluster using simulated Chandra observations tailored to examine gas sloshing effects. Their findings indicate that the density fluctuations observed beyond the central 60 kpc align closely with those expected from sloshing phenomena alone. Furthermore, their analysis translated these surface brightness fluctuations into estimates of turbulent heating rates. Remarkably, across the annuli examined beyond 60 kpc, the estimated heating rates were found to either match or exceed the cooling rates. 
 
 Similarly, the residual image of Abell 2151 exhibits a possible spiral feature that extends radially to approximately 81 kpc. To determine whether these interactions are driving gas turbulence and to assess their relative contributions to the total variance, an `effective' equation of state can be used to study the nature of gas perturbations. This equation, which illustrates the correlation between density fluctuations and temperature fluctuations, can be derived by measuring fluctuations in both the soft energy band ($0.5 - 3.5$ keV) and the hard energy band ($3.5 - 7$ keV). The equation is defined as follows \citep{osti_1256639}:

\begin{equation}
    \frac{\delta T}{T} = (\xi_i - 1) \frac{\delta \rho}{\rho},
\end{equation}

where $\xi_i$ represents the effective adiabatic index. Given that emissivity is more strongly temperature-dependent at higher energies, surface brightness fluctuations in the hard band are used to characterize temperature fluctuations. The gas perturbations are classified into three types according to their origins. Isothermal perturbations ($\xi_{i}=1$), manifesting as variations in thermal gas density at a constant temperature, are typically associated with bubbles of relativistic plasma. Isobaric fluctuations ($\xi_{i}=0$), often induced by slow gas displacements such as the motion of galaxies and mergers, are sensitive to entropy changes while maintaining pressure equilibrium with the surrounding atmosphere. Adiabatic fluctuations ($\xi_i=5/3$), produced by shock fronts, sound waves, and other mildly transonic disturbances, are characterized by constant gas entropy. 

Isobaric perturbations, predominant in the core of Abell 2029 on scales of $60 - 90$ kpc, constitute more than 70 percent of the total variation, measured by \citep{zhuravleva2016nature}, suggesting that slow motions of gas or gas cooling might significantly contribute to these gas disturbances in the core. They also estimated that approximately 80 percent of perturbations in the core are isobaric at scales of $\sim 8 - 70$ kpc in the core of Perseus, where a similar spiral feature is observed. It indicates that the central AGN activity is not the only driver of gas perturbations in the ICM. However, the existing exposure of Chandra observations does not provide sufficient counts in the hard band, preventing us from determining the nature of gas perturbations in the remaining four clusters. 

While we have not observed significant X-ray cavities in the five clusters studied, it is possible that the dissipation timescale of perturbations produced by feedback processes is long. This implies that we could have a mix of both AGN feedback and other mechanisms, such as gas sloshing, contributing to the generation of turbulence. Future observations and more detailed analyses will be required to determine the relative contributions of these processes.

On the other hand, mild turbulence has been suggested as a potential trigger for thermal instabilities \citep{voit2018role,gaspari2018shaken}. However, the lack of cooling products in Abell 2029 and Abell 2151 could imply that the mergers occurring in these clusters may not generate sufficient turbulence levels to induce thermally unstable cooling.

The variation of metallicity can also cause emissivity variation, which might further lead to an overestimation of density fluctuations. The nuclear metallicity of Abell 2029 lies near the solar value and then drops to $\sim 0.4~ Z_{\odot}$ at 100 kpc. \citet{kirkpatrick2015hot} found evidence for enhanced metallicity compared to the ambient value with an amplitude of $\rm \sim 0.1~Z_\odot$ in several regions along the radio axis. While the magnitude of this signal is comparable to the measurement errors, its systematic presence across multiple regions is notable. This apparent excess spans several tens of kpc in altitude. Nevertheless, at this level, metallicity variations are unlikely to significantly contribute to the observed emissivity/density variations in Abell 2029. Furthermore, in the other four clusters, significant metallicity variance was also not observed. The density fluctuations due to cooling don't trace velocity, however, it can be neglected because the eddy turnover timescale $\rm t_{eddy}$ is typically shorter than cooling timescale $\rm t_{cool}$. 
 
The mechanisms contributing to the observed emissivity/density fluctuation spectra in these smooth atmospheres remain elusive. Future research, potentially involving more simulations of clusters lacking radio bubbles yet with short central cooling times will aim to quantify the effects of halo disturbances and galaxy motions. More insight will come from direct high-resolution X-ray spectroscopy including the X-ray Imaging and Spectroscopy Mission (XRISM), which has recently published its initial results featuring a non-dispersive resolution of $\sim$ 5 eV.


\subsection{Shape of velocity spectra \& Mass Dependence}
\label{sec: slope}

For pure hydrodynamic turbulence, the energy spectrum $E(k_1)$, where $k_1 = 2\pi/l = 2\pi k$, depends solely on the wavenumber $k_1$ and the mean density-normalized dissipation rate is $\rm \epsilon = Q_{turb}/\rho_0$ within the inertial scale range. This relationship is expressed as $E(k_1) = C_K\epsilon^{2/3}k_1^{-5/3}$, with $C_K$ representing the universal Kolmogorov constant. In the case of isotropic turbulence, the mean square component of the velocity field is uniform in all directions, and as such, $V_{1,k} = V_k/3 = [2k_1E(k_1)/3]^{1/2}$ can be approximated as $\sim k^{-1/3}$.  

Pure turbulence in this setting may be unrealistic as the effects of gravity and bulk motions should lead to departures from the Kolmogorov spectrum. The errors in Figure~\ref{fig:ps_density_fluctuations} are consistent with a range of slopes, including $\propto k^{-1/3}$. So we cannot exclude pure turbulence. However, the surface brightness fluctuations may not be caused entirely by turbulent motions. Therefore, the measurements presented in Figure~\ref{fig:velocity_ps} should be considered upper limits on turbulent velocities. Assuming that gas perturbations are caused only by turbulence, the heating rates would be in the same order as the cooling rates, as shown in Figure~\ref{fig: heating_cooling_rate}. 

 Several interesting theoretical possibilities exist to explain slope variance. They include turbulent energy injected at multiple spatial scales, and a time-dependent evolution of turbulent atmospheric motions \citep[]{chen2023empirical}. The steepening may indicate partial dissipation of certain modes or suppression of power on small scales. For instance, magnetic fields can steepen the kinetic power spectrum when magnetic stresses suppress the kinetic energy cascade \citep[]{bambic2018suppression}. However, when including magnetohydrodynamic (MHD) effects, plasma effects should also be considered, in which the pressure anisotropy causes the turbulent plasma to resist changes in magnetic-field strength, thereby preserving the spectrum slope~\citep{squire2019magneto}.

It is also worth noting that the $\Delta$-variance method used in X-ray surface brightness fluctuation analysis may introduce slope bias \citep{arevalo2012mexican,romero2024forecasting}. Furthermore, the depth of the Chandra exposures limits the attainable accuracy. These considerations restrict our measurements to a relatively narrow scale range and prevent us from accurately constraining the spectral slopes.

\begin{figure}
    \centering
    \includegraphics[width=\columnwidth,height=7.0cm]{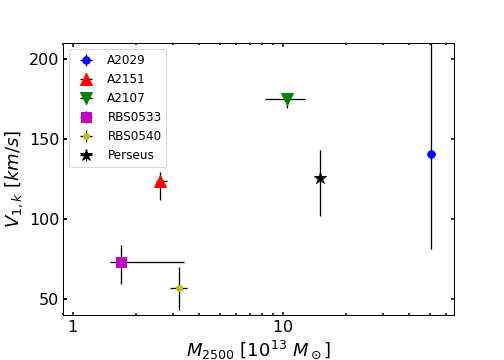}
    \caption{\small{$M_{2500}$ versus one-component velocity $V_{1,k}$ of smooth-core clusters and Perseus cluster. The markers present the mean velocity amplitudes over the range of scales probed in each target.}}
    \label{fig: mass_velocity}
    \end{figure}

Finally, we examine the trend between cluster mass and turbulent velocity shown in Figure~\ref{fig: mass_velocity}. Total mass estimates are found by \citet{martz2020thermally} using the Navarro–Frenk–White (NFW) potential model. Their masses range between $\rm 10^{13}~M_\odot$ and $\rm 10^{14}~M_\odot$ as detailed in the column (b) of Table~\ref{tab:velocity_spectra}. The vertical error bars reflect the $1\sigma$ uncertainties in the turbulent velocity; the horizontal error bars reflect the uncertainties in mass profile fitting.  No significant correlation between mass and turbulent velocity is found.

%% file: Cpt6_Conclusion.tex

\section{Conclusions}
\label{sec: conclusion}
In this work, we studied gas perturbations in the atmospheres of five cool-core clusters, which have short central cooling times and low entropy but show little evidence of cold gas and bubbles inflated by AGN feedback. Assuming that the detected density fluctuations reflect underlying random gas motions, the amplitudes of gas density fluctuations estimated from X-ray surface brightness fluctuations are used to derive the velocity spectra of gas motions. We estimated the dissipation heating rates and further explored the impact of turbulent dissipation in reheating and stabilizing the atmospheres of the structureless clusters. The uncertainties due to the limited exposure times, the choice of the underlying SB models, and the deviation from the Kolmogorov slope are also examined. The key conclusions drawn from our research are as follows:

\begin{itemize}
    \item The characteristic amplitude of density fluctuations is $\sim 10$ per cent at the scale of the cool core. Amplitudes rise to $\sim 15$ per cent at the scale of $\sim 216$ kpc in Abell 2029 and $\sim 17$ per cent at the scale of $\sim 35$ kpc in Abell 2151. Despite the absence of structures associated with central AGN activity, our results are comparable to the gas density fluctuations measured in the core of the Perseus cluster, where a large fraction of the core area is occupied by bubbles, shocks, and spirals. 

\item The measured power spectra are consistent with a broad range of slopes, making it impossible to confirm or rule out alignment with the Kolmogorov turbulence.

    \item The velocity amplitudes lie below $\rm 100\ km\ s^{-1}$ in RBS0540 and RBS0533 but rise to $\rm \sim 178\ km\ s^{-1}$ at scales of $\sim 30$ kpc in Abell 2107 and $\rm \sim 211\ km\ s^{-1}$ at scales of $\sim 220$ kpc in Abell 2029. These velocity amplitudes are consistent with turbulent velocities expected in hot atmospheres. 
    
    \item No relationship is found between cluster mass, $\rm M_{2500}$, and the velocity amplitudes.
    
     \item The turbulent heating rates implied by our measurements are of the same order as the radiative cooling rates. Our results suggest that atmospheric sloshing and perhaps turbulent motion may aid radio jets in stabilizing atmospheric cooling.

\end{itemize}

%% file: Acknowledgements.tex

\section*{Acknowledgements}
\label{sec:ack}

 M.L expresses appreciation to Prathamesh Tamhane and Connor Martz for their valuable discussions and for generously sharing their spectral-fitting scripts. B.R.M. acknowledges the support from the Natural Sciences and Engineering Research Council of Canada and the Canadian Space Agency Space Science Enhancement Program. I.Z acknowledges partial support from NASA through Chandra Award Number AR4-25012X, issued by the Chandra X-ray Observatory Center, which is operated by the Smithsonian Astrophysical Observatory for and on behalf of NASA under contract NAS8-03060.

%% file: data_availability.tex
\section*{DATA AVAILABILITY}
\label{sec:daa}
This study made use of data from the public Chandra Data Archive and software provided by the Chandra X-ray Center (CXC) through the CIAO package. The data products will
be shared on reasonable request to the corresponding author.